\documentclass[12pt]{iopart}
\usepackage{iopams}

\usepackage{amssymb}

\usepackage{graphicx}
\usepackage{fancybox,color}

\usepackage[latin1]{inputenc} % Umalaute

\def\xnewpage{} %{\clearpage} %{\newpage}

\newlength{\myboxL}

\def\black{\color{black}}

\renewcommand{\Re}{\mbox{Re}}

\def\intpi{\frac1{2\pi}\int_0^{2\pi}}
\def\intpia{\frac{\a}{\pi}\int_{-\pi/2\a}^{\pi/2\a}}
\def\Ocal#1{{\mathcal O}\left(#1\right)}
\def\H{{\cal H}}
\def\F{{\cal F}}
\def\M{{\cal M}}
\def\hc{{\rm{h.c.}}} %\def\hc{\text{h.c.}}
\def\d{\partial}
\def\h{\hat}

\def\s{{}^\dagger}
\def\v#1{|#1\rangle}

\def\la{\lambda}
\def\al{\alpha}

\def\vac{|0\rangle}
\def\cav{\langle 0|}

\def\D{\Delta}
\def\a{\ell}
\def\J{J}

\def\ra{\rightarrow}

\newcommand{\ket}[1]{\left|#1\right>}

\newcommand{\braket}[2]{\left<#1|#2\right>}
\newcommand{\nn}{\nonumber\\}

\newcommand{\bea}{\begin{eqnarray}}
\newcommand{\ea}{\end{eqnarray}}
\newcommand{\eea}{\end{eqnarray}}

\def\eqref#1{(\ref{#1})}
\def\text#1{{\rm{#1}}}

\begin{document}
%%%%%%%%%%%%%%%%%%%%%%%%%%%%%%%%%%%%%%%%%%%%%%%%%%%%%%%%%%%%%%%%%%%%%%%%%%%%%%%

\title[Optical lattice quantum simulator for QED in strong external fields]
{Optical lattice quantum simulator for QED in strong external fields:
spontaneous pair creation and the {Sauter-Schwinger} effect
}

\author{N Szpak and R Sch{\"u}tzhold}

\address{Fakult{\"a}t f{\"u}r Physik, Universit{\"a}t Duisburg-Essen,
Duisburg, Germany}

\eads{nikodem.szpak@uni-due.de, ralf.schuetzhold@uni-due.de}

\date{\today}

\begin{abstract}
{
Spontaneous creation of electron-positron pairs out of the vacuum due
to a strong electric field is a spectacular manifestation of the
relativistic energy-momentum relation for the Dirac fermions.
This fundamental prediction of Quantum Electrodynamics (QED) has not yet
been confirmed experimentally as the generation of a sufficiently strong
electric field extending over a large enough space-time volume still
presents a challenge.
Surprisingly, distant areas of physics may help us to circumvent this
difficulty.
In condensed matter and solid state physics (areas commonly considered
as low energy physics), one usually deals with quasi-particles instead
of real electrons and positrons.
Since their mass gap can often be freely tuned, it is much easier to
create these light quasi-particles by an analogue of the Sauter-Schwinger
effect.
This motivates our proposal of a quantum simulator in which excitations
of ultra-cold atoms moving in a bichromatic optical lattice represent
particles and antiparticles (holes) satisfying a discretized version of
the Dirac equation together with fermionic anti-commutation relations.
Using the language of second quantization, we are able to construct an
analogue of the spontaneous pair creation which can be realized in an
(almost) table-top experiment.
}
\end{abstract}

\maketitle
%%%%%%%%%%%%%%%%%%%%%%%%%%%%%%%%%%%%%%%%%%%%%%%%%%%%%%%%%%%%%%%%%%%%%%%%%%%%%%%
\section{Introduction}

Spontaneous creation of electron-positron (fermion-antifermion) pairs from
vacuum under specific external conditions is a direct manifestation of the
relativistic energy-momentum relation for the Dirac particles.
The most prominent realization of this effect is creation and separation
of an electron ($e^-$) and a positron ($e^+$) in presence of a strong
electric field, derivable as a phenomenon within Quantum Electrodynamics
(QED).
The adiabatic character of the spontaneous pair creation allows for the
interpretation in which the particle is slowly pulled from the otherwise
unobservable \textit{Dirac sea} while the hole in the \textit{sea} appears
as an antiparticle.
Both come into being on the cost of the external fields.
Unfortunately, electrons and positrons, the lightest fermions satisfying
the Dirac equation \cite{Dirac}, still defend themselves from being exposed
in that way.
{
Generation of a strong enough electric field which is able to deliver the
minimal energy $2\, m_e c^2 = 1.022$ MeV needed to create a pair from vacuum
still appears to be an experimental challenge.
A natural
}
source of a strong and localized electric field, the atomic nucleus,
would need to carry a charge of at least $+173\,$e
(slightly depending on its predicted size, see e.g.~\cite{Nuclei}) what is
about $50$ unit charges above the heaviest (and unstable) nuclei which have
ever been observed in the laboratory.
In the early 1980s, there have been serious experimental attempts \cite{EPOS}
to collide beams of fully ionized uranium atoms $U^{92+}$ or similar ions
{
in order to create a sufficiently long-lived charge concentration of around
$+184\,$e but they were not successful in this regard.
Recent developments -- e.g., in the field of strong lasers \cite{ELI} or
the current extension of GSI in Darmstadt --
have again renewed interest in spontaneous pair creation.
}
%
%\mybox{cite review/sth new?}
%
However, we are clearly not yet in the position of creating electric fields
of sufficient strength.

Quite surprisingly, help {may} come from distant areas of physics:
{
condensed matter and solid state physics
}
-- areas commonly considered as low energy physics.
Since the energy scale is {determined by the mass of the particles
under consideration, the electron mass} can set a too high barrier for
electrons and positrons while the {analogous gap} can be much lower
for light quasi-particles whose masses can be tuned in experiments.
This motivates our proposal of a quantum simulator in which excitations of
ultra-cold atoms moving in a regular optical lattice will represent particles
and antiparticles (holes) satisfying a discretized version of the Dirac
equation together with fermionic anti-commutation relations.
Applying the language of second quantization, we construct an analogue of
the spontaneous pair creation which can be realized in an (almost)
table-top experiment.

To additionally motivate the need of a quantum simulator, we mention some
{open} problems still present in theory and experiment related to
supercritical fields of QED.
The simplest setting in which the spontaneous pair creation should occur is
the case of a constant electric field $E$, well known in the literature as
the \textit{Schwinger effect} or \textit{Sauter-Schwinger effect}
\cite{Schwinger, Sauter, Heisenberg+Euler}.
For nonzero values of {the electric field} $E>0$ one should observe
spontaneously generated {pairs} of particles and antiparticles with
probability {(per unit time and volume)} given by
\begin{equation}
\label{P-Schwinger}
P_{e^+e^-}
\sim
\exp\left\{-\pi\,\frac{c^3}{\hbar}\,\frac{M^2}{qE}\right\}
=
\exp\left\{-\pi\,\frac{E_{\text{S}}}{E}\right\}
\,,
\end{equation}
where ${E_{\text{S}}}=M^2c^3/(\hbar q)$ is the critical field strength
determined by the elementary charge $q$ and the mass $M$ of an electron
(or positron).
{Besides the aforementioned experimental difficulties, the above}
expression for $P_{e^+e^-}$ is non-perturbative in $q E$ and does not
permit any expansion in the field strength $E$ nor in the coupling constant
(or charge) $q$, e.g. via a finite set of Feynman diagrams.
{Thus,}
apart from the constant field case, only very simple field configurations,
where the electric field either depends on time $E(t)$ or on one spatial
coordinate such as $E(x)$,
{have been treated analytically so far} \cite{calculations}.
Consequently, our theoretical understanding of various aspects of this
effect under general conditions is still quite limited.
For example, recently it has been found that the occurrence of two
different frequency scales in a time-dependent field $E(t)$ can induce
drastic changes in the (momentum dependent) pair creation probability
\cite{catalysis,Stokes}.
Moreover, the impact of interactions between the electron and the
positron of the created pair, as well as between them and other
electrons/positrons {or photons is still not fully understood.}
This ignorance is unsatisfactory not only from a theory point of view
but also in view of planned experiments with field strengths
not too far below the critical field strength ${E_{\text{S}}}$ and thus
capable of probing this effect experimentally \cite{ELI}.

The proposed quantum simulator will reproduce the quantum many-particle
Hamiltonian describing electrons and positrons in strong electric fields
and should thereby reproduce the \textit{Sauter-Schwinger effect}.
This will facilitate investigation of space-time dependent electric fields
such as $E(t,x)$ and also provide new insight into the role of interactions
which may be incorporated into the simulator.

It should be stressed here that our proposal goes beyond the simulation
of the (classical or first-quantized) Dirac equation on the single-particle
level, see, e.g.,
\cite{Witthaut+Weitz-DoublePeriodicOptPot, RS+Unruh-SlowLight, Dirac-photons,
Dirac+Zitterbewegung-photons, QSim-Dirac, QSim-Dirac-TrappedIon,
Schaetz-QSim-Dirac-TrappedIon},
but aims at the full quantum many-particle Hamiltonian.
A correct description of many-body effects such as particle-hole creation
(including the impact of interactions) requires creation and annihilation
operators in second quantization.
There are some proposals for the second-quantized Dirac Hamiltonian
\cite{QSim-Dirac-HexLattice, Dirac+Interaction-OptLat,
MasslessDirac-SqLattice, Dirac-StaggeredMagnField,
Goldman+Lewenstein-Dirac-by-SU2, Lewenstein-Dirac-CurvedST}
but they consider scenarios which are more involved than the set-up
discussed here and aim at different models and effects.
Similarly, the recent observation of Klein tunneling in graphene
\cite{Graphene-Dirac} deals with massless Dirac particles --
but the mass gap is crucial for the non-perturbative
{Sauter-Schwinger}  effect, cf.~Eq.~(\ref{P-Schwinger}).
{Furthermore, graphene offers far less flexibility than optical
lattices regarding the experimental options for changing the relevant
parameters or single-site and single-particle addressability, etc.}

%%%%%%%%%%%%%%%%%%%%%%%%%%%%%%%%%%%%%%%%%%%%%%%%%%%%%%%%%%%%%%%%%%%%%%%%%%%%%%%
\section{Spontaneous pair creation
in supercritical external fields} \label{sec:SpPCr}
%\subsection{The Dirac equation}

We consider the Dirac equation \cite{Dirac} describing electrons/positrons
propagating in an electromagnetic vector potential $A_\mu$
which are described by the spinor wave-function $\Psi$
($\hbar=c=1$)
\begin{equation}
\gamma^\mu (i\d_\mu-qA_\mu) \Psi - M\, \Psi = 0
\,.
\end{equation}
For simplicity, we consider 1+1 dimensions ($\mu=0,1$) where the
Dirac matrices $\gamma^\mu$ satisfying the Clifford algebra
$\{\gamma^\mu,\gamma^\nu\}=2\eta^{\mu\nu}$
can be represented by Pauli matrices
$\gamma^0=\sigma_3$ and $\gamma^1=-i\sigma_1$.
Furthermore, we can choose the gauge $qA_0=\Phi$ and $A_1=0$
(because in one spatial dimension, there is no magnetic field).
In one spatial dimension, there is also no spin, hence the wave function
has only two components $\Psi=(\Psi^1,\Psi^2)$.
As a result, the Dirac equation simplifies to
\begin{equation}
i\d_t\Psi(t,x)
=
\mathcal{H}\,\Psi(t,x) = (-i \sigma_2 \d_x  + M \sigma_3 + \Phi) \Psi(t,x).
\end{equation}
%
%
%%%%%%%%%%%%%%%%%%%%%%%%%%%%%%%%%%%%%%%%%%%%%%%%%%%%%%%%%%%%%%%%%%%%%%%%%%%%
%\subsection{Spectrum and supercriticality}
%
If $\Phi(x)$ is negative and vanishes at infinity sufficiently fast the
spectrum of {the Dirac Hamiltonian} $\mathcal H$ consists of two
continua $(-\infty, -M] \cup [M,\infty)$ and a discrete set of bound states
$E_n$ lying in the gap $(-M,M)$.
%\begin{minipage}{0.45\linewidth}
%\end{minipage}
%\begin{minipage}{0.50\linewidth}
%\end{minipage}

\begin{figure}[h]
  \begin{center}
     \includegraphics[width=0.45\linewidth]{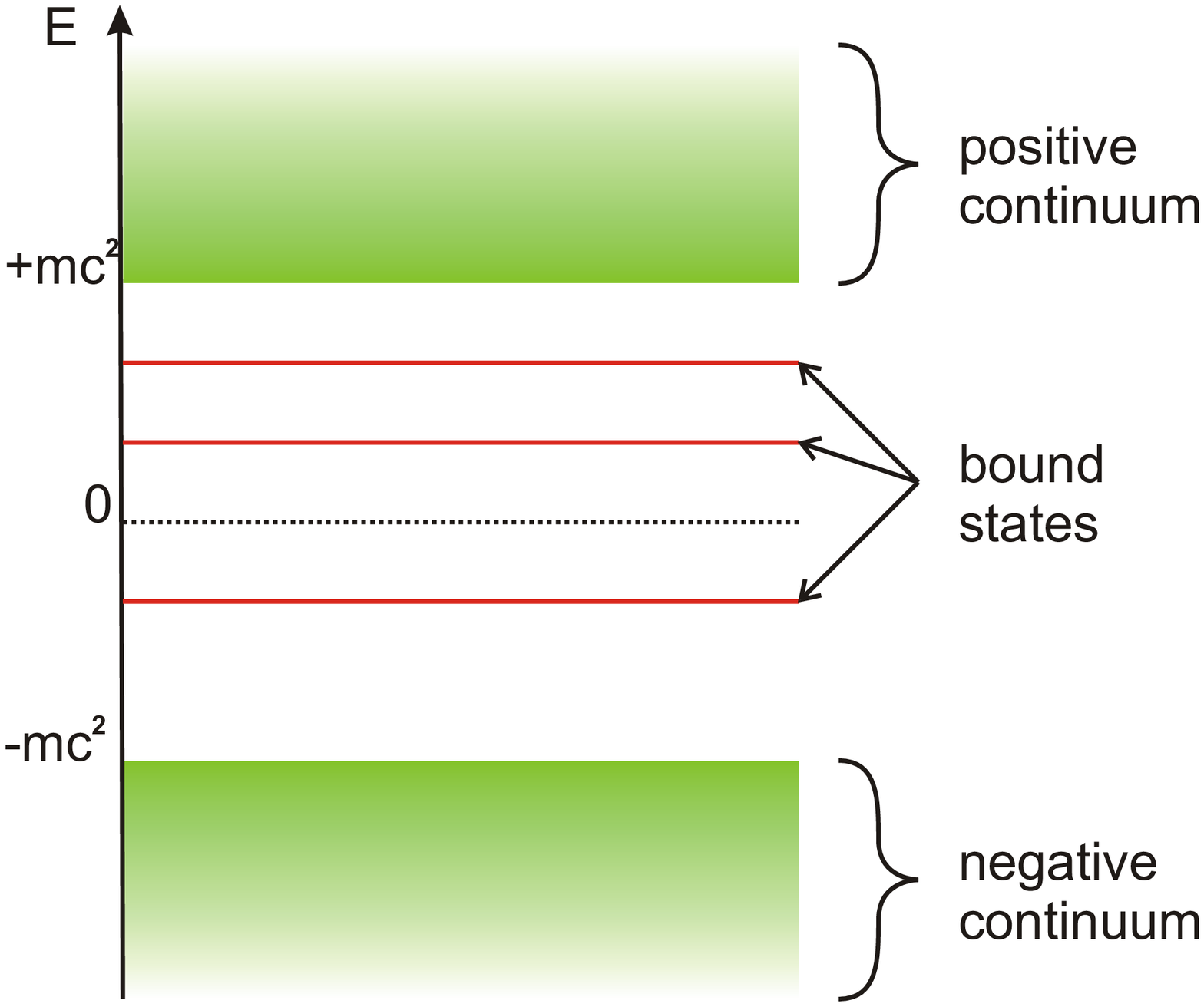}
     \hfill
     \includegraphics[width=0.5\linewidth]{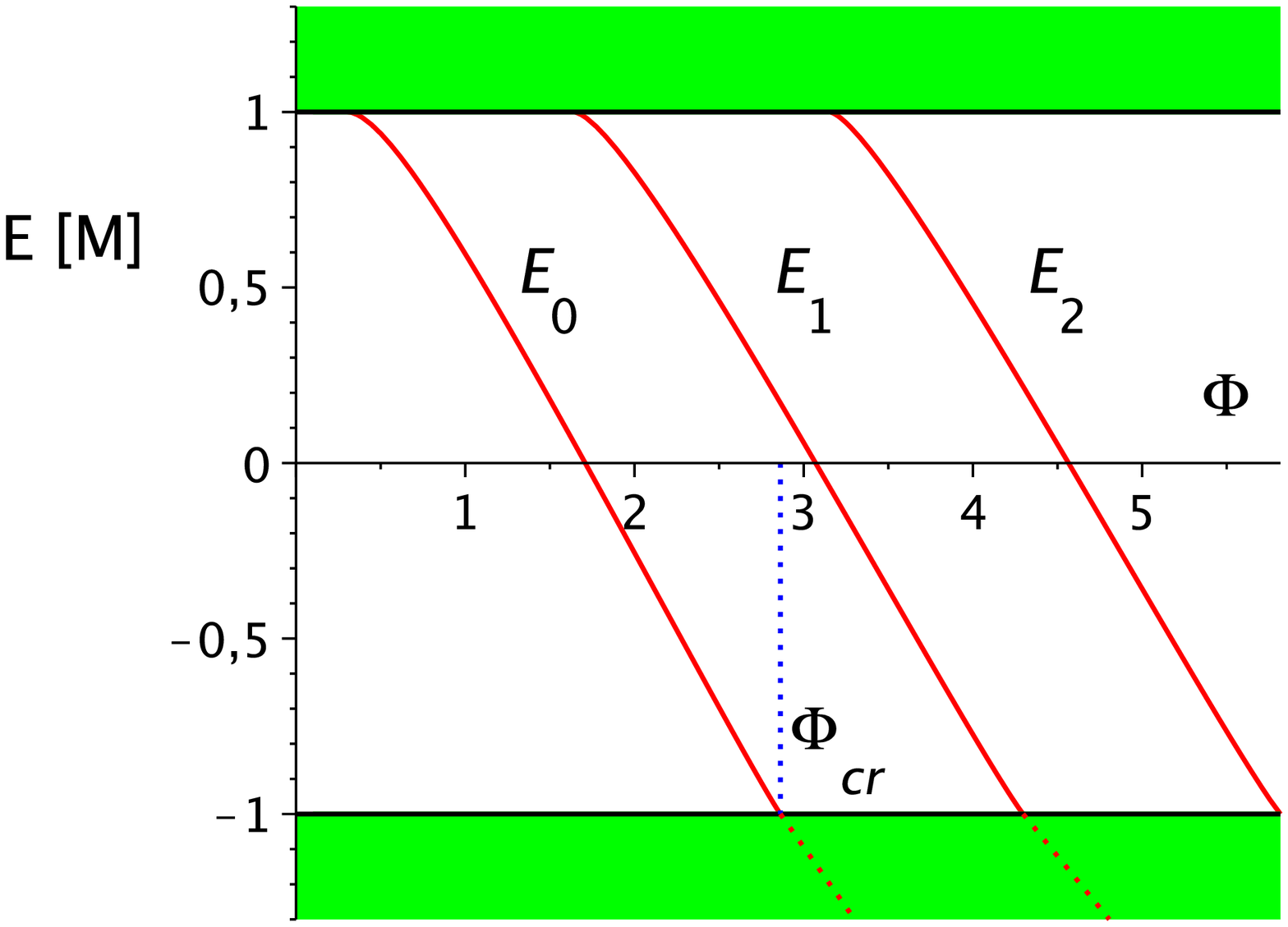}
  \end{center}
\caption{Typical spectrum of a Dirac Hamiltonian (left) and its dependence
on the strength of an attractive potential $\Phi$ (right).
At the critical value $\Phi=\Phi_\text{cr}$ the lowest bound-state $E_0$
(red solid line) turns to a resonance in the negative continuum
(red dotted line).
Next bound states $E_1, E_2, ...$ follow for larger values of $\Phi$.}
  \label{fig:Dirac-spectrum}
\end{figure}

The bound-state energies $E_n$ depend continuously on the parameters of the
potential.
In particular, as the strength of the negative potential $|\Phi|$ increases
each $E_n \searrow -M$.
Already at a finite value $\Phi_\text{cr}$, called \textit{critical}, the
lowest lying bound state $E_0$ reaches the negative continuous spectrum
associated with the interpretation of antiparticles,
i.e. $E_0|_{\Phi=\Phi_\text{cr}}=-M$.
For \textit{supercritical} strength of the potential $|\Phi|>|\Phi_\text{cr}|$
the bound state -- corresponding to a real pole in the resolvent of
$\mathcal H$ (or in the scattering operator) -- turns to a resonance
(complex pole) with $\Re(E_0)<-M$ (see Fig. \ref{fig:Dirac-spectrum}).

%%%%%%%%%%%%%%%%%%%%%%%%%%%%%%%%%%%%%%%%%%%%%%%%%%%%%%%%%%%%%%%%%%%%%%%%%%%%
%\subsection{Spontaneous pair creation}

\begin{figure}[h]
  \begin{center}
  \includegraphics[width=0.6\linewidth]{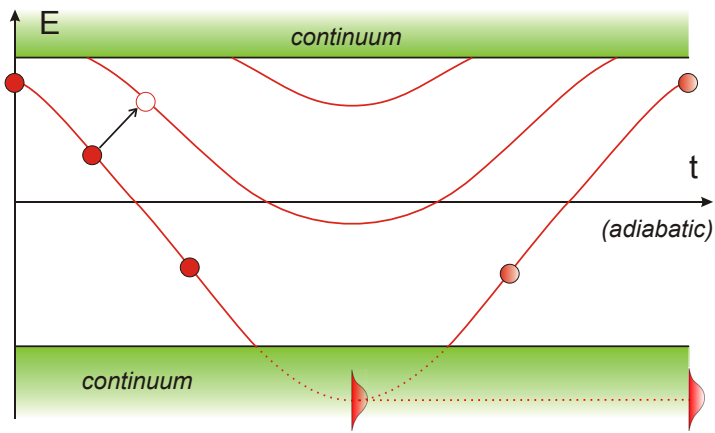}
  \end{center}
\caption{Spectrum of the Dirac Hamiltonian $\H(t)$ in presence of a slowly
varying potential $\Phi(t)$.
In the middle, a supercritical phase: the lowest bound-state
(red solid line) enters the negative continuum and turns to a resonance
(red dotted line).
\label{fig:E_t}}
\end{figure}

Imagine now a time-dependent process in which $\Phi(t)$ is slowly varied
between the sub- and supercritical regimes as in Fig. \ref{fig:E_t}.
In agreement with the adiabatic theorem, in the subcritical phase, the
{quantum state} of the system follows the eigenstate in which it is
initially prepared.
As the supercritical phase begins {the gap closes and} the
adiabatic theorem breaks down \cite{Nen80, Nen87}.
The system follows then a resonance which is spectrally represented by a wave
packet with position and width varying in time.
Such wave packets inevitably decay in the lower continuum, trapping a big
part of the wave function.
Therefore, during the switch-off phase, when the potential $\Phi(t)$ becomes
subcritical again, only a small part of the wave function follows the
re-appearing eigenstate.
Mathematically, there exists a non-vanishing matrix  element of the
scattering operator between the positive ($+$) and negative ($-$) continuum,
$S^{+-}\neq 0$, which tends to one in the ``adiabatic limit''.
In order to avoid interpretational problems  we need to leave the
one-particle picture at this stage and switch to the many-particle
{description.
In the language of second quantization, the discussed process is described
by the scattering operator $\h S$ acting in a Fock space and it can be determined from the one-particle counterpart \cite{Scharf}. It involves \textit{dynamical} and \textit{spontaneous} creation of pairs as well as annihilation and scattering of already present particles and antiparticles.
%
%\red
%Since the Dirac field operator $\hat\Psi$ can be written as a linear combination of the annihilation operators $\hat a_I$ for electrons and the creation operators $\hat b_J^\dagger$ for positrons, the most general form of $\hat S$ for linear Dirac fields in time-dependent external potentials assumes the following form
%
%\bea
%\hat S=
%\exp\left(
%S^{+-}_{IJ}\hat a_I^\dagger\hat b_J^\dagger
%+
%S^{-+}_{IJ}\hat a_I\hat b_J
%+
%S^{++}_{IJ}\hat a_I^\dagger\hat a_J
%+
%S^{--}_{IJ}\hat b_I^\dagger\hat b_J
%\right)
%\,. \mybox{\tra $S^{+-} S^{-+}$}
%\ea
%
%\blue
%\begin{equation}
%\fl
%\hat S= C
%\exp\left(A^{+-}_{IJ}\hat a_I^\dagger\hat b_J^\dagger\right)
%\exp\left(B^{-+}_{IJ}\hat a_I\hat b_J\right)
%:\exp\left(C^{++}_{IJ}\hat a_I^\dagger\hat a_J\right) :\
%:\exp\left(D^{--}_{IJ}\hat b_I^\dagger\hat b_J\right):
%\end{equation}
%\black
%The first term $S^{+-}_{IJ}$ describes \textit{dynamical} or \textit{spontaneous} creation of particle-antiparticle pairs and the second term $S^{-+}_{IJ}$ their annihilation while the last two terms only scatter already present particles and antiparticles, leaving the vacuum vector $\ket\Omega$ untouched.
%
In the ``adiabatic limit'', the \textit{dynamical} pair production goes
to zero such that only the \textit{spontaneous} process remains.
%
%\red
%In the scenario discussed above (see Fig.~\ref{fig:E_t}), it is sufficient to consider two modes only (one stemming form the bound state in the positive continuum \mybox{???} and one corresponding to the decaying resonance in the negative continuum)
%
%\bea
%\hat S\to
%\exp\left(
%S^{+-}\left[\hat a^\dagger\hat b^\dagger+\hat a\hat b\right]
%\right)
%\cdot \h S_0
%\,, \qquad\mybox{?}
%\ea
%
%\red
%where $\h S_0$ scatters within the two continua
%$\h S_0 \ket{\Omega} = \ket\Omega$.
%
%Starting in the vacuum state, we may restrict our analysis to the sub-space of vanishing total charge, where we find
%
%\bea
%\hat S\to
%\cos(S^{+-})\h S_0+
%\left[\hat a^\dagger\hat b^\dagger+\hat a\hat b\right]
%\sin(S^{+-})\h S_0
%\,. \qquad\mybox{?}
%\ea
%
%In the ``adiabatic limit'', we obtain $S^{+-}=\pi/2$, i.e., one pair is created from vacuum with unit probability.
%
%$\h S =
%\exp\left(S_{+-}\,\h a\s_{+} \h a\s_{-} - S_{+-}^*\,\h a_{+} \h a_{-}\right)
%\cdot \h S_0$,
%where the exponential in front of $\h S_0$ creates particle-antiparticle
%pairs while $\h S_0$ only scatters already present particles and
%antiparticles, leaving the vacuum vector $\ket\Omega$ untouched
%$\h S_0 \ket{\Omega} = \ket\Omega$.
%%
%In the ``adiabatic limit'', the \textit{dynamical} pair production goes
%to zero while one pair, corresponding to
%$\h S \to \h a\s_+ \h a\s_- \cdot \h S_0$
%is still created from vacuum.
%
Therefore, for processes (as described above) starting from a vacuum state $|\Omega\rangle $ and running through a
supercritical phase we obtain
\cite{Nen80, NS-PhD}
\begin{equation}
|\Omega\rangle \quad \longrightarrow \quad \h S |\Omega\rangle
=
\h a\s \h b\s |\Omega\rangle
\end{equation}
while for subcritical processes $|\Omega\rangle \ra |\Omega\rangle$,
in agreement with the adiabatic theorem.}
This phenomenon is called \textit{spontaneous pair creation}, as opposed to
the \textit{dynamical pair creation}, since  it is related to the
\textit{spontaneous} decay of a time-dependent ground state $\v{\Omega(t)}$
(in the so called \textit{Furry picture}) during the supercritical phase
\cite{RafelskiMuellerGreiner, NS-PhD}.

%%%%%%%%%%%%%%%%%%%%%%%%%%%%%%%%%%%%%%%%%%%%%%%%%%%%%%%%%%%%%%%%%%%%%%%%%%%%
\xnewpage
\section{Discretized quantum Dirac field}

In this section we make the first step towards the quantum simulator of a
quantum Dirac field in optical lattices and discretize the theory by
introducing a regular lattice in space. We will argue that
the phenomena of strong external fields like the supercriticality and
spontaneous pair creation discussed above will survive this operation.

%\subsection{Dirac field}

The Hamiltonian for the classical Dirac field reads
\begin{equation}
H = \int dx\,{\Psi}\s (-i \sigma_2 \d_x  + M \sigma_3 + \Phi) \Psi
\,.
\end{equation}
We introduce a regular grid (lattice) $x_n = n \cdot \a$ with a positive
grid (lattice) constant $\a$ and integers $n\in\mathbb Z$.
The discretization of the wave function $\Psi_n(t):=\sqrt{\a}\,\Psi(t,x_n)$,
defined now at the grid points $x_n$, gives rise to a discretized derivative
$\sqrt{\a}\,\d_x \Psi(t,x_n) \to [\Psi_{n+1}(t)-\Psi_{n-1}(t)]/(2\a)$
and to a discretized potential $\Phi_n := \Phi(x_n)$.
Finally, replacing the $x$-integral by a sum, we obtain
\begin{equation}
\label{continuum}
H_d = \sum_n \Psi\s_n \left[-\frac{i\sigma_2}{2\a} (\Psi_{n+1}-\Psi_{n-1})
+ M\sigma_3 \Psi_n + \Phi_n\Psi_n\right].
\end{equation}
In order to obtain the quantum many-body Hamiltonian, we quantize the
discretized Dirac field operators via the fermionic anti-commutation
relations
\begin{equation}
\label{anti-commutation}
\{\h\Psi_n^\alpha(t),[\h\Psi_m^\beta(t)]\s\}=
\delta_{nm} \delta^{\alpha\beta}
\,,\qquad
\{\h\Psi_n^\alpha(t),\h\Psi_m^\beta(t)\}=0
\;.
\end{equation}
Substituting %$\hat\Psi_n=(\hat a_n,\hat b_n)$, i.e.,
$\hat\Psi_n^1=\hat a_n$ and $\hat\Psi_n^2=\hat b_n$ the discretized
many-particle Hamiltonian obtains the form
\begin{eqnarray}
\fl
\h H
&=&
\frac1{2\a} \sum_n
\left[ \h b^\dagger_{n+1} \h a_n - \h b^\dagger_n \h a_{n+1} + \hc \right]
+\sum_n \left[(\Phi_n+M)\h a\s_n \h a_n + (\Phi_n-M) \h b\s_n \h b_n \right]
\,.
\end{eqnarray}
The first term describes jumping between the neighboring grid points while the
remaining two terms can be treated as a combination of external potentials.
Due to the specific form of the jumping, the lattice splits into two
disconnected sub-lattices: (A) containing $\h a_{2n}$ and $\h b_{2n+1}$
and (B) containing $\h a_{2n+1}$ and $\h b_{2n}$ with integers $n$.
Since the two sub-lattices behave basically in the same way,
it is sufficient to consider only one of them, say A.
With a re-definition of the local phases via
$\h a_{2n} \to (-1)^n \h a_{2n}$ and $\h b_{2n+1} \to (-1)^{n+1} \h b_{2n+1}$,
we obtain (for sub-lattice A)
\begin{eqnarray}
\label{H-lattice-a-b}
\h H
&=&
-\frac1{2\a} \sum_n
\left[ \h b\s_{2n+1} \h a_{2n} + \h b\s_{2n-1} \h a_{2n} + \hc \right]
+
\\
&&
+\sum_n
\left[
(\Phi_n+M) \h a\s_{2n} \h a_{2n} +(\Phi_n-M) \h b\s_{2n+1} \h b_{2n+1}
\right]
\,.
\nonumber
\end{eqnarray}
Identifying $ \h c_{2n} = \h a_{2n} $ and $ \h c_{2n+1} = \h b_{2n+1}$,
this takes the form of the well known Fermi-Hubbard Hamiltonian for a
one-dimensional lattice
\begin{equation}
\label{Fermi-Hubbard0}
\h H = - \frac{J}{2} \sum_{n}
\left[\h c\s_{n+1} \h c_n + \h c\s_{n} \h c_{n+1}\right] +
\sum_n V_n \h c\s_n \h c_n
\,,
\end{equation}
with hopping rate $J=1/\a$ and on-site potentials $V_n=\Phi_n + (-1)^n M$.
This Hamiltonian will be the starting point for the design of the optical
lattice analogy.

Alternatively, using a more abstract language of modern quantum field theory,
the Hamiltonian $\h H$, being an operator {acting on the} Fock space
$\F$, can be directly obtained by implementation of the discretized
single-particle Hamiltonian %$\mathcal{H}_d$
%The action of the discretized one-particle Hamiltonian is
\begin{equation}
\mathcal{H}_{d} \Psi_n
=
-\frac{i}{2\a} \sigma_2 (\Psi_{n+1}-\Psi_{n-1}) +
M\sigma_3 \Psi_n + \Phi_n \Psi_n
\end{equation}
acting in the discretized Hilbert space $\mathfrak{H}_d=(L^2(\mathbb{Z}))^2$
({which is the} discretization of $\mathfrak{H}=L^2(\mathbb{R})^2$)
as a self-adjoint operator in the Fock space $\F$ according to
%\begin{multline}
\begin{eqnarray}
%\begin{split}
\eqalign{
  \h H &= \sum_n \h\Psi^*(\mathcal{H} f_n) \h\Psi(f_n)
       =
\frac1{2\a} \sum_n \left[ \h\Psi^{2\,\dagger}_{n+1}
\h\Psi^1_n - \h\Psi^2_n\s \h\Psi^1_{n+1} + \hc \right]\\
       &+\sum_n \left[(\Phi_n+M)\h\Psi_n^1\s \h\Psi_n^1 +
(\Phi_n-M)\h\Psi_n^2\s \h\Psi_n^2 \right].}
%\end{split}
\end{eqnarray}
%\end{multline}
where the  second quantized discretized Dirac field operator
%$\h \Psi^\al_n(t)$
%
\bea
\h \Psi^\al_n := \h \Psi^\al(f_n) = \int dx\, \h \Psi^\al(x) f_n^*(x)
\ea
%
%$\h \Psi^\al_n := \h \Psi^\al(f_n) = \int \Psi^\al(x) \overline{f_n(x)} dx$
satisfies the above anti-commutation relations
%$ \{\h \Psi_n(t),\h \Psi\s_l(t)\} = \delta_{n,l} ...$
and the orthonormal set of basis functions $f_n$ spans the discretized
Hilbert space $\mathfrak{H}_d$.
(Here, no charge conjugation or renormalization is needed as we will
physically deal with finite systems only.)

%%%%%%%%%%%%%%%%%%%%%%%%%%%%%%%%%%%%%%%%%%%%%%%%%%%%%%%%%%%%%%%%%%%%%%%%%%%%
%%%%%%
%
\subsection{Spectrum}
The free part $\hat H_0$ of this Hamiltonian, i.e., {without} the
external potential $\Phi_n=0$, can be explicitly diagonalized.
Performing a discrete Fourier transform on the lattice
%\begin{eqnarray}
%   \h a_{2n} &= \frac{1}{\pi} \int_0^\pi e^{2ikp} \h a(p) dp, &
%   \h b_{2n+1} &= \frac{1}{\pi} \int_0^\pi e^{i(2n+1)p} \h b(p) dp
%\end{eqnarray}
\begin{equation} \label{a-b-Fourier}
\h a(p) := \sum_n e^{-i2n\a p} \h a_{2n},\quad \h b(p)
:=
\sum_n e^{-i(2n+1)\a p} \h b_{2n+1},
\end{equation}
for $p\in\left[-\pi/2\a,+\pi/2\a\right)$,
where the anti-commutation relations (\ref{anti-commutation}) imply
\begin{equation}
  \{\h a(p), \h a(q)\s\} = \frac{\pi}{\a} \delta(p-q),
\qquad \{\h a(p)\s, \h a(q)\s\} = \{\h a(p), \h a(q)\} = 0,
\end{equation}
we obtain
\begin{equation}
\label{H0-a-b-cos}
\fl  \h H_0 = \intpia dp
\left[ M \left( \h a(p)\s \h a(p) - \h b(p)\s \h b(p) \right)
+ \frac1{\a}\cos(\a p)
\left( \h a(p)\s \h b(p) + \h b(p)\s \h a(p)\right)\right]
\,.
\end{equation}
This Hamiltonian can be diagonalized via a unitary transformation mixing
the two types of operators
\begin{equation} \label{diag-a(p)}
  %\left(\begin{array}{c}
  \left(\begin{array}{l}
    \h A(p) \\ \h B(p)
  \end{array}\right)
  %\end{array}\right)
   = U(p)
  %\left(\begin{array}{c}
  \left(\begin{array}{c}
    \h a(p) \\ \h b(p)
  \end{array}\right)
  %\end{array}\right)
\end{equation}
with the explicit form
\begin{equation} \label{U(p)}
  U(p) = \frac{1}{\sqrt{2E}} %\left(\begin{array}{c}{}
     \left(\begin{array}{cc}
     \sqrt{E+M} & \sqrt{E-M} \\
     -\sqrt{E-M} & \sqrt{E+M}
  \end{array}\right)
  %\end{array}\right)
\end{equation}
what leads to
\begin{equation} \label{H0-a-b-sqrt}
\h H_0 = \intpia dp\,
E(p) \left[ \h A(p)\s \h A(p) - \h B(p)\s \h B(p)\right]
\end{equation}
%\begin{equation}
%\label{H0-a-b-sqrt}
%\h H_0 = \int%\limits^{+\pi/(2\a)}_{-\pi/(2\a)}
%dp\, E_p
%\left[\hat A_p^\dagger\hat A_p - \hat B_p^\dagger\hat B_p\right]
%%\,.
%\end{equation}
with the energy-momentum relation
\begin{eqnarray}
\label{E_p}
E(p)=\sqrt{M^2+\frac1{\a^2}\cos^2(\a p)}
\,.
\end{eqnarray}
Due to two effective types of fermionic excitations,
$\h A(p)$ and $\h B(p)$, which enter the Hamiltonian with opposite energy
signs, we obtain two symmetric energy bands separated by a gap of $2M$.
Each approximates the relativistic energy-momentum relation at the edge
of the Brillouin zone, for $p\approx \pm\pi/(2\a)$.
In order to obtain a positive Hamiltonian, we can perform the usual
re-definition $\hat B(p)^\dagger\leftrightarrow\hat B(p)$ which
corresponds to changing the vacuum state by filling all $\hat B(p)$
states with fermions.
This is analogous to the \textit{Dirac sea} picture in full quantum
electrodynamics.
In terms of this analogy, $\hat A(p)^\dagger$ or $\hat A(p)$ create or
annihilate an electron whereas $\hat B(p)$ or $\hat B(p)^\dagger$
create or annihilate a positron.
An additional potential $\Phi_n$, if sufficiently localized in space,
will not modify this spectrum but may introduce
bound states (isolated eigenvalues) with energies lying in the gap
\cite{ImpurityStates}.

%%%%%%%%%%%%%%%%%%%%%%%%%%%%%%%%%%%%%%%%%%%%%%%%%%%%%%%%%%%%%%%
\subsection{Supercritical potential}

As an example for demonstration of supercriticality in the discretized
system we consider the attractive Woods-Saxon
potential\footnote{However, the discussed phenomena are generic and do
not depend on the details of the potential.}
\begin{equation} \label{WS-pot}
  \Phi(x) = - \frac{W}{1+e^{a(|x|-L)}}, \qquad W, a, L > 0,
\end{equation}
for which the one-dimensional Dirac equation is analytically solvable in
terms of hypergeometric functions.
The bound-state energies $E_n$ are determined by the equation
\begin{equation} \label{WS-bst-cond}
  \frac{B(-2g,g+s-\lambda)^2}{B(2g,-g+s+\lambda)^2} =
e^{4gaL} \frac{(s-g)^2-\lambda^2}{(s+g)^2-\lambda^2},\qquad
  B(x,y):=\frac{\Gamma(x)\Gamma(y)}{\Gamma(x+y)}
\end{equation}
with $s:=\sqrt{M^2-E^2}/a, g:=i\sqrt{(E+W)^2-M^2}/a, \lambda:=iW/a$
and depend continuously on the parameters of the potential
\cite{Kennedy:Woods-Saxon}.

\begin{figure}[h]
  \begin{minipage}{0.75\linewidth}
     \includegraphics[width=1\linewidth]{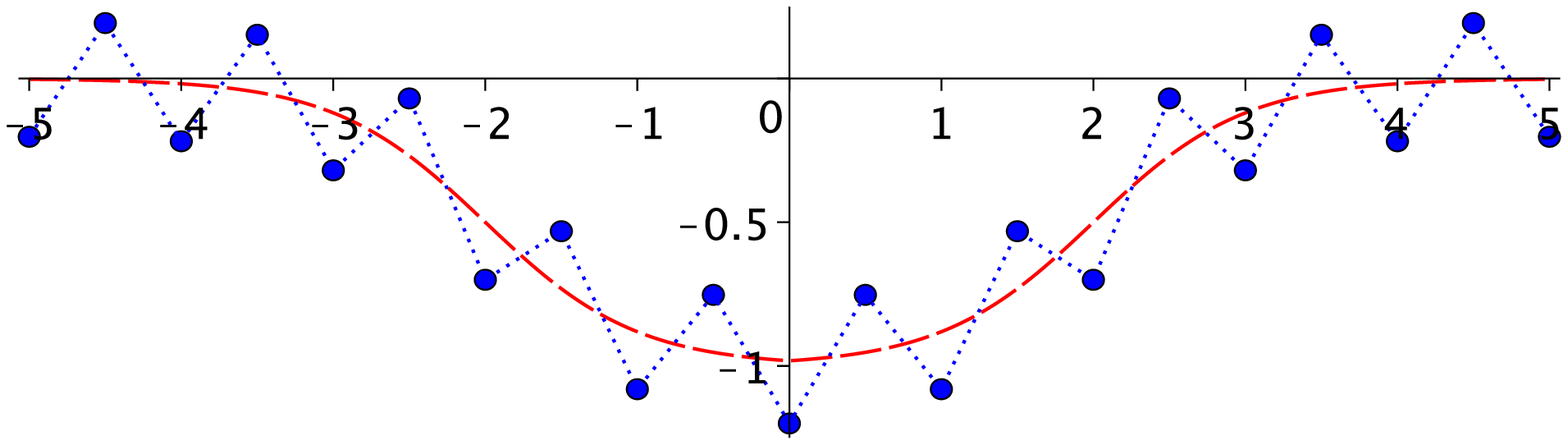}
  \end{minipage}
  \hfill
  \begin{minipage}{0.20\linewidth}
     \includegraphics[width=1\linewidth]{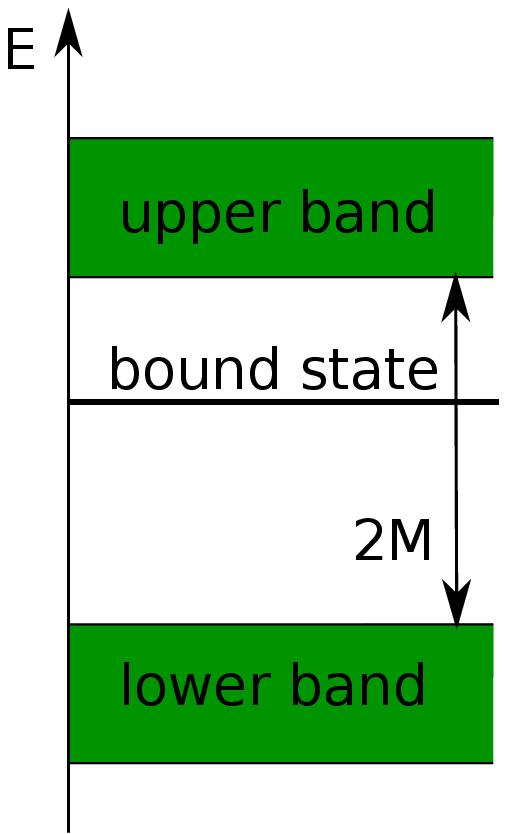}
  \end{minipage}
  \caption{Left: The continuous Woods-Saxon potential $\Phi(x)$ (red dashed line)
and the discretized effective potential $\Phi_n + (-1)^n M$
(blue dots, connected with dotted line for visualization only).
  Right: The corresponding energy spectrum. (Not to scale.)}
  \label{fig:WS+discr}
\end{figure}
\begin{figure}[h!]
  \begin{center}
     \begin{minipage}{0.55\linewidth}
       \includegraphics[width=1\linewidth]{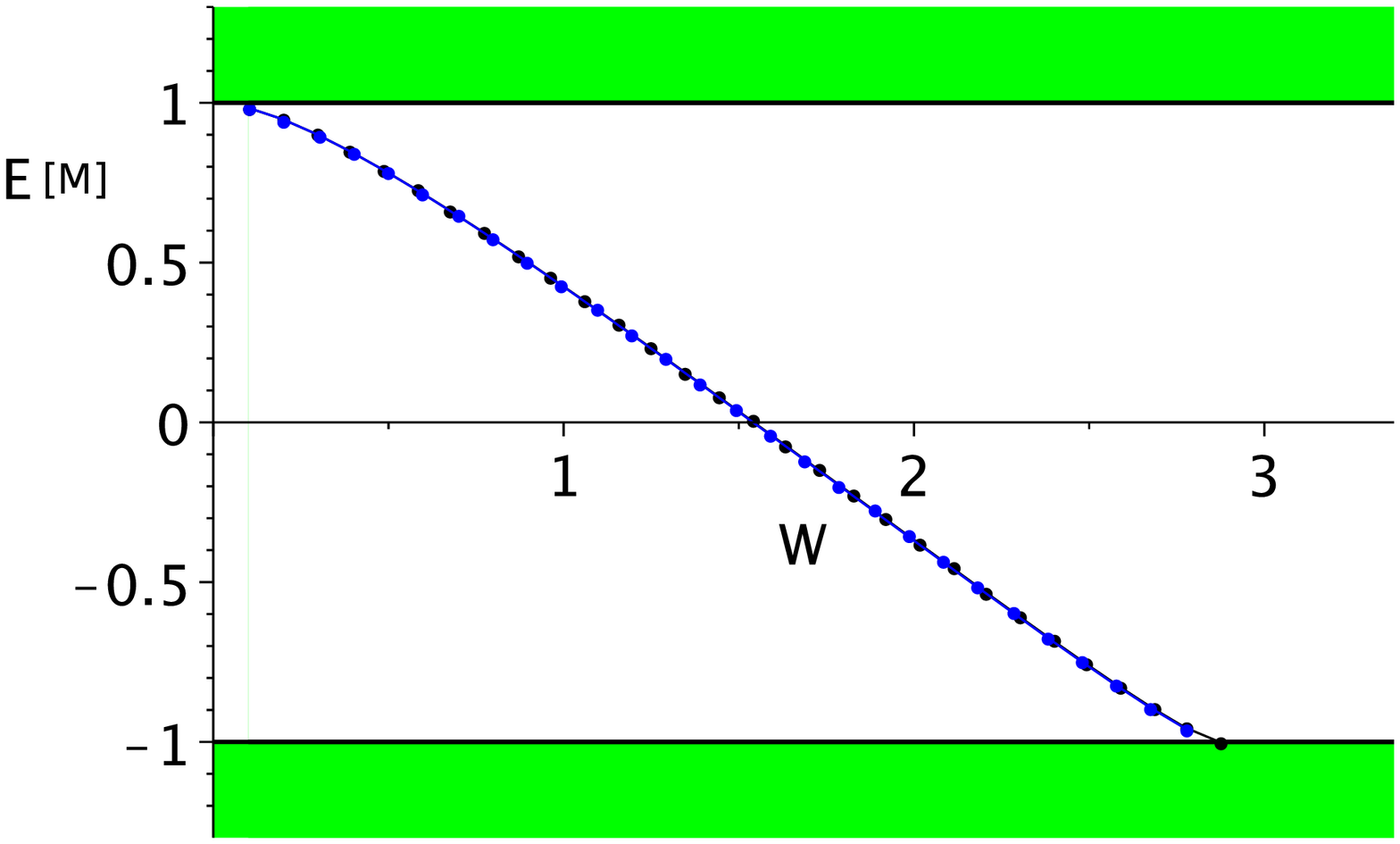}
     \end{minipage}
     \hfill
     \begin{minipage}{0.4\linewidth}
       \includegraphics[width=1\linewidth]{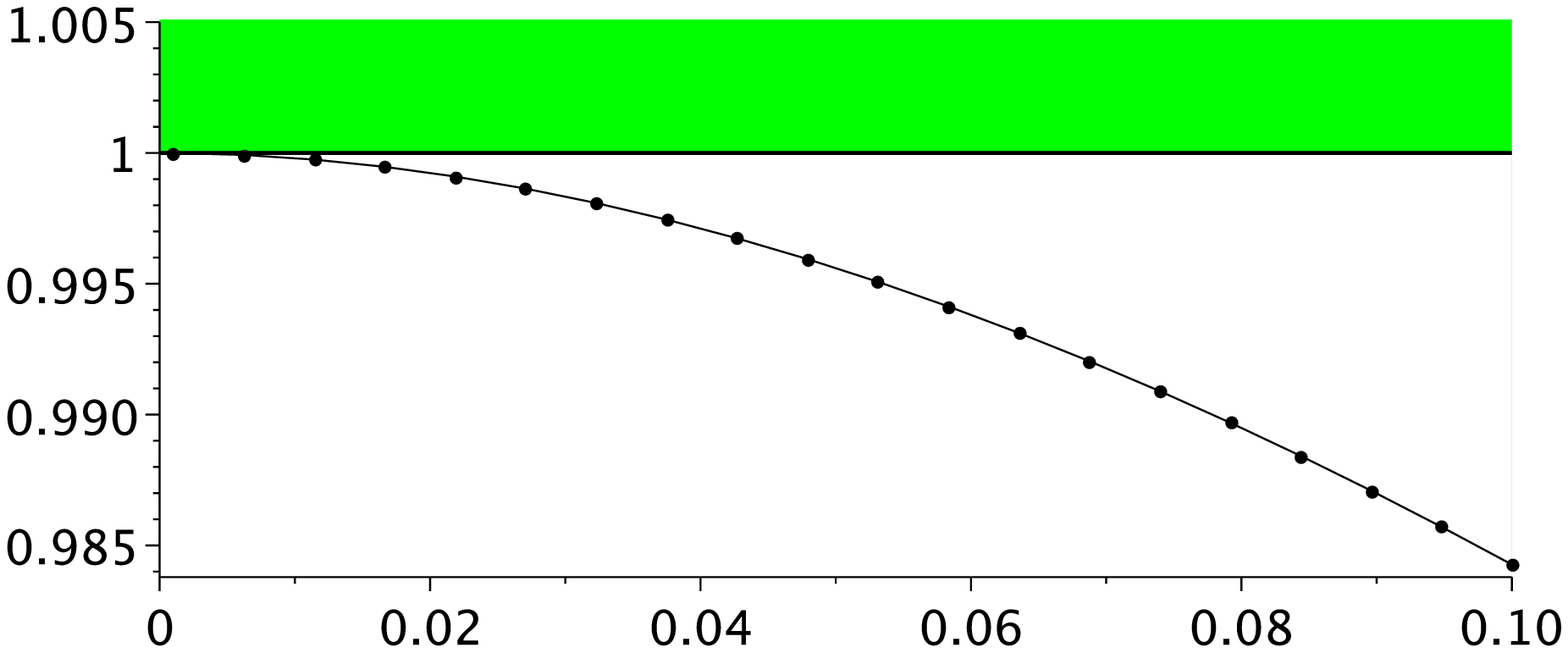} \vfill
       \includegraphics[width=1\linewidth]{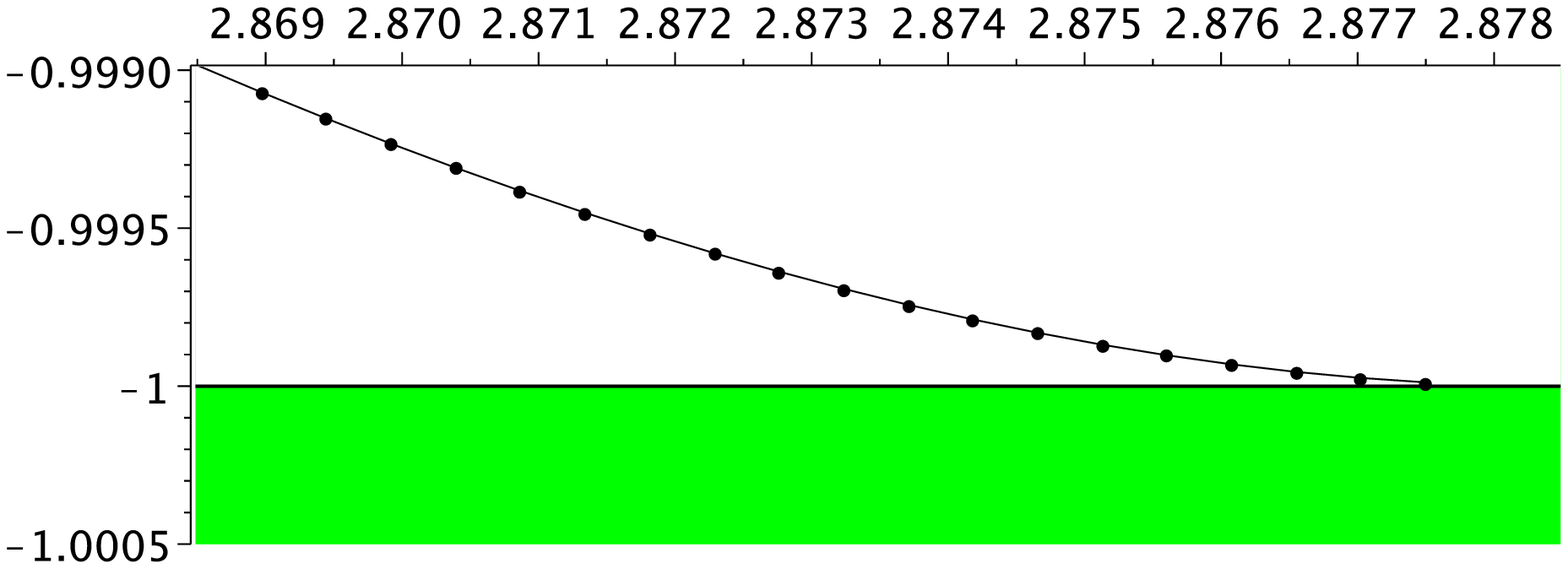}
     \end{minipage}
  \end{center}
  \caption{Comparison of the bound state energy $E$ as a function of the
potential strength $W$ for the continuous (black dots and line) and
discretized (blue dots and line) Dirac equation with Woods-Saxon potential
\eqref{WS-pot} ($a=10, L=1$). On the right, the parabolic
approach, $E(W) \approx \pm M + C_\pm (W-W_\pm)^2$, to the upper and
lower continuum.}
  \label{fig:WS+discr-bst}
\end{figure}

Below, we compare the spectra of the continuous and the discretized Dirac
equations with the Woods-Saxon potential.
In the latter case, the discretized Woods-Saxon potential is defined by
$\Phi_n := \Phi(x_n)$ with $x_n = \a\cdot n$ (see Fig. \ref{fig:WS+discr}).
For both cases we calculate numerically the lowest-lying bound state $E_0$
as a function of the parameter $W$ ($a$ and $L$ fixed) which is a monotone
function with $dE_0/dW<0$ as long as $-M < E_0 < M$.
The dependence of the bound state energies $\tilde E_0(W)$ for the Hubbard
Hamiltonian \eqref{Fermi-Hubbard0} on the parameter $W$ is qualitatively
the same and quantitatively in very good agreement with the curve $E_0(W)$
obtained as solution of \eqref{WS-bst-cond} in the continuous case.
At almost the same critical value $W_\text{cr} \approx 2.878$ both bound
states disappear from the spectrum and turn into complex
resonances\footnote{In one dimension this process is slightly more
complicated {than} the well-known diving of the bound state into the
continuum in 3 dimensions. Here, the bound state first turns at the
threshold $E=-M$ into an anti-bound state ($E>-M$ and $\sqrt{M^2-E^2}$
changes sign) and moves slightly up to eventually turn again and  dive
in the negative continuum $E<-M$ as a resonance, cf. \cite{Dreizler}.}.
The curves are compared in Fig. \ref{fig:WS+discr-bst}.
\begin{figure}[h]
  \begin{center}
     \includegraphics[width=0.8\linewidth]{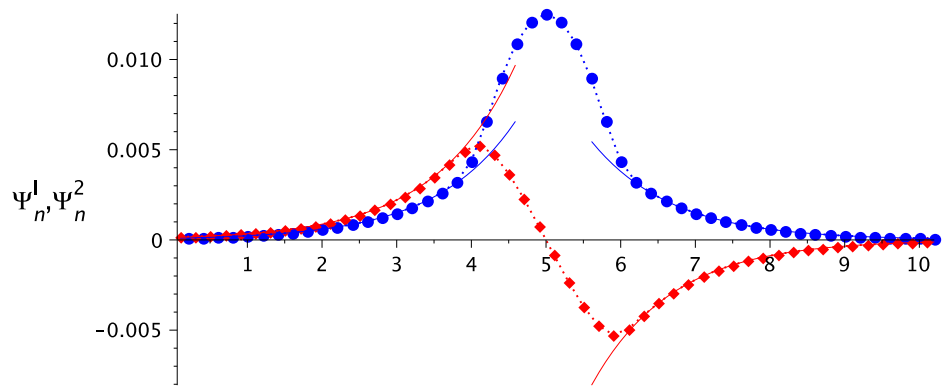}
  \end{center}
  \caption{Typical profile of a bound state (here $E=-0.37 M$).
Plotted are two components of the wave function $\Psi^1_n, \Psi^2_n$
(blue dots and red diamonds) against $x=n\cdot d$, supplemented by
exponential asymptotes (blue and red solid lines).
One function has always one zero and the other has no zeros
like in the continuous Dirac equation.}
\end{figure}

%%%%%%%%%%%%%%%%%%%%%%%%%%%%%%%%%%%%%%%%%%%%%%%%%%%%%%%%%%%%%%%%%%%%%%%%
\xnewpage
\section{Quantum simulator}
\subsection{Ultra-cold atoms in optical lattices}

The main goal of this work is to propose a physical quantum system, composed
of ultra-cold atoms moving in a specially designed periodic potential,
which can effectively be described with a Fermi-Hubbard-type Hamiltonian
of the form (\ref{Fermi-Hubbard0}) and pseudo-relativistic dispersion
relation (\ref{E_p}) approximating the relativistic formula $E^2=M^2+p^2$
for energies around the Fermi-level. Moreover, excitations of the ground
state should behave like particles and antiparticles and obey the Fermi
statistics.

Ultra-cold atoms loaded into optical lattices are conveniently described
by effective discrete Hubbard-type Hamiltonians \cite{HubbardModels-OptLat}.
That kind of approximation is based on a construction of a set of
orthonormal wave functions  $\psi_n$ (\textit{Wannier functions})
localized around the local minima of the potential $W(x)$,
giving rise to a regular grid of \textit{sites}, and on the assumption
that the single-particle Hamiltonian $\H$ is approximately
tri-diagonal in that basis, i.e. $\braket{\psi_n}{\H|\psi_m}\approx 0$
for $|n-m|>1$. In consequence, the many-body Hamiltonian can be written
in the form (\ref{Fermi-Hubbard0}) with
$J_n = \braket{\psi_n}{\H| \psi_{n+1}}$ and
$V_n = \braket{\psi_n}{\H| \psi_n}$.
There is a deeper connection between that approximation, in which only the
lowest energy band is taken into account in the construction of the Wannier
functions, and a spatial discretization of the theory in which the
discretization step (equal to the period of the potential) introduces
a natural cut-off in energies. In the latter approximation, the
coefficients $J_n$ and $V_n$ correspond to the discretized kinetic
(Laplacian) and potential terms in the Hamiltonian.
In both approaches the energy spectrum is reduced to a single band.

\subsection{Bi-chromatic optical lattice}

\begin{figure}[h]
\begin{center}
\includegraphics[width=0.5\linewidth]{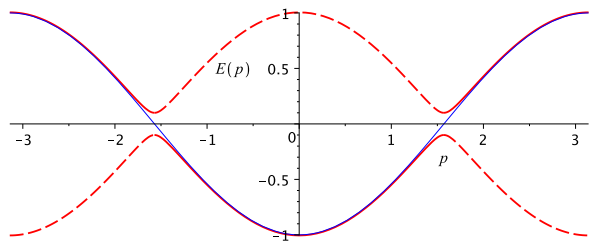}
\end{center}
\caption{Sketch of the typical dispersion relation
%(in units of $J$ and $1/a$) for $\Delta W=0$
for a periodic potential (solid blue line) and
%small $\Delta W>0$
perturbed doubly-periodic one
(dashed red curves).}
\label{fig:band-split}
\end{figure}

It turns out that
the emergence of a pseudo-relativistic dispersion relation, as in
Eq.~(\ref{E_p}), is a quite universal phenomenon, see also \cite{RS+NS, Bands}.
\black
Imagine, we start with a periodic potential in one spatial dimension and
introduce a small perturbation which breaks the original periodicity and
is only periodic with the double period.
This implies that the Brillouin zone $[-\pi/(2\a), \pi/(2\a)]$ shrinks by a factor of two and that
the lowest band splits into two sub-bands. %of half the width in $p$-space.
Since, at the same time, the perturbation is small the energy-momentum
relation $E(p)$ at any given momentum $p$ can only change by a small
amount.
In consequence, the perturbation will induce significant changes only
in the vicinity of the momenta $p_0=\pm \pi/(2\a)$,
i.e. edges of the shrinked Brillouin zone,
%where the original dispersion relation $E_0(p)$ \mybox{went trough zero} $E_0(p_0)=0$.
at which it generates a small gap in the spectrum
which separates the two branches of $E(p)$,
see Figure \ref{fig:band-split}.
For small perturbations, this gap will be proportional to the amplitude
of the perturbation \cite{RS+NS, Bands}.
Altogether, we reproduce the pseudo-relativistic dispersion relation
in the vicinity of those points $p_0$.
%
%
%\bigskip
%
%As we discuss it in more detail in \cite{RS+NS, Bands} the lowest band
%splits into two when the periodic potential gets perturbed by a function
%which has a double period. The origin of this phenomenon lies in  shrinking
%of the Brillouin zone by factor 2 (in one dimension) caused by doubling of
%the potential period.
%Since, at the same time, the perturbation is small the energy-momentum
%relation can change only slightly within the corresponding zone.
%In consequence, \blue two branches of $E(p)$ appear with a small gap in
%the middle whose width is proportional to the amplitude of the
%perturbation (as long as it stays small).
%The result is presented in Figure \ref{fig:band-split}.
%%%%%%%%%%%%%%%%%%%%%%%%%%%%%%%%%%%%%%%%%%%%%%%%%%%%%%%%%%%%%%%%%%%%%%%%%%%%%%%
%%%%%%%%%%%%%%%%%%%%%%%%%%%%%%%%%%%%%%%%%%%%%%%%%%%%%%%%%%%%%%%%%%%%%%%%%%%%%%%

The conditions for a quantum simulator formulated above can be satisfied,
in a good approximation,
with ultra-cold fermionic atoms loaded into a one-dimensional optical lattice
with the doubly-periodic potential
\begin{eqnarray}
 W(x) = W_0 \sin^2(2kx) + \Delta W \sin^2(kx)
\,,
\end{eqnarray}
\begin{figure}[h]
  \begin{center}
  \begin{minipage}{0.7\linewidth}
    \includegraphics[width=1\linewidth,height=0.2\textheight]{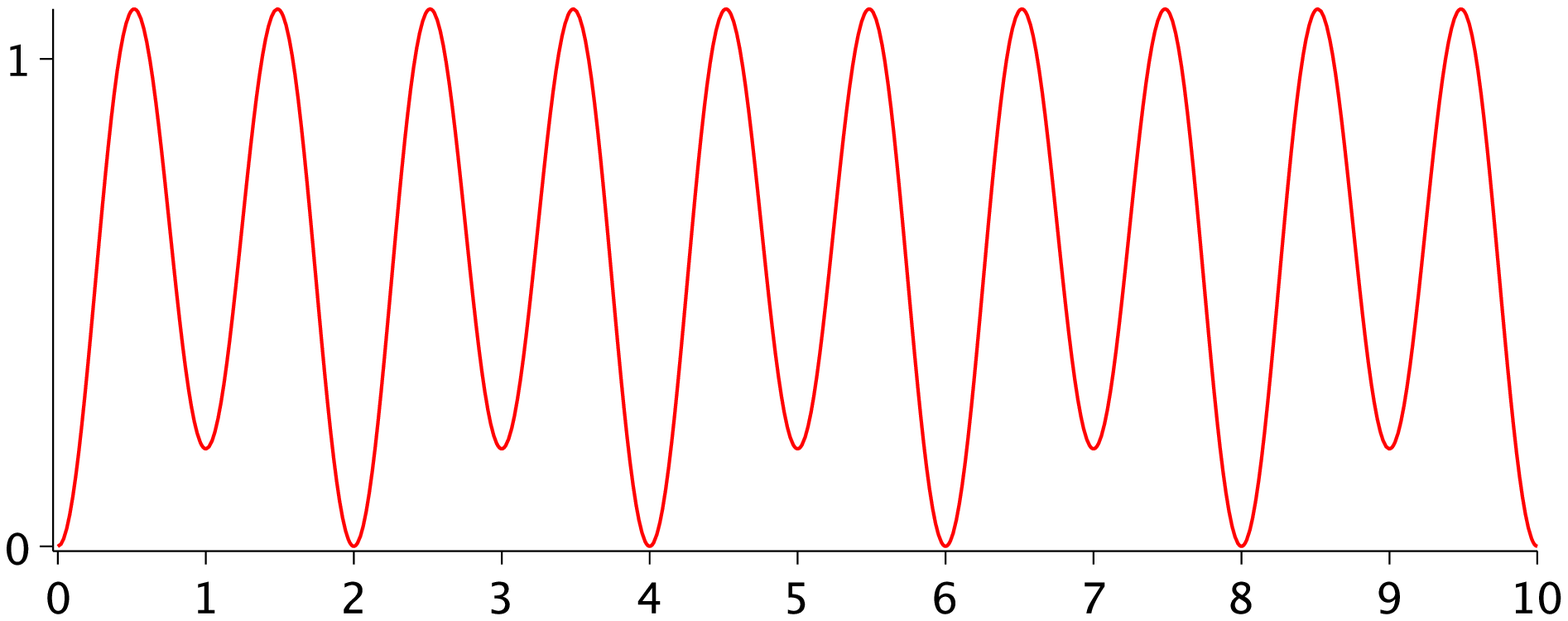}
  \end{minipage}
  \begin{minipage}{0.20\linewidth}
     \includegraphics[width=1\linewidth,height=0.2\textheight]{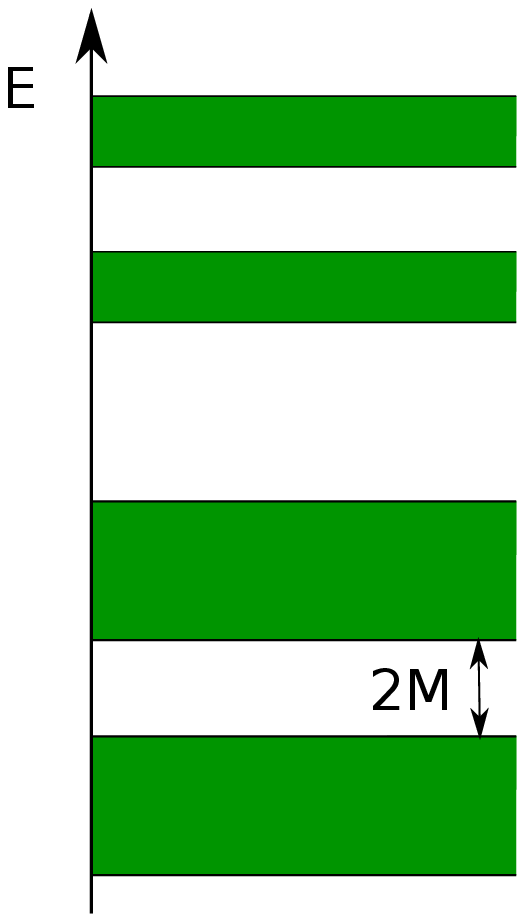}
  \end{minipage}
  \end{center}
  \caption{Left: Doubly-periodic potential $W(x)$.
  Right: The corresponding energy spectrum with the gap $2M \approx \D W$ between the lowest two bands. (Not to scale.)}\label{fig:potential}
\end{figure}
where $k=\pi/(2\a)$, by taking $W_0\gg\Delta W$ (see Fig. \ref{fig:potential}).
Potentials of that form can be obtained by superposition of two
lattice-generating standing laser waves with different frequencies.
Similar settings have already been obtained experimentally
\cite{Weitz-DoublePeriodicOptPot}.

{Unfortunately,} it is not possible to find a closed analytic formula
for the energy-momentum dependence in that potential.
However, by applying a version of the WKB method for periodic potentials
\cite{Balazs-WKB} to the doubly-periodic case we were able to derive
\cite{Bands} a spectral condition from which an approximate dispersion
relation can be calculated analytically. That condition reads
\begin{equation}
  \cos^2(\Phi/2) = (1-T)\, \sin^2(\Delta \Phi) + T \, \cos^2(\a p).
\end{equation}
where $T(E)$ is the WKB-transmission coefficient through a single potential
barrier [around a maximum of $W(x)$], $\Phi:=\Phi_1+\Phi_2$ and
$\Delta \Phi=\Phi_1-\Phi_2$ with the WKB phases
\begin{equation}
  \Phi_i(E):=\int_{y_i}^{z_i} \sqrt{E-W(x)}\,dx
\end{equation}
calculated between two consecutive turning points $y_i, z_i$ corresponding to the
same potential minimum for which  $W(y_i)=W(z_i)=E$ and $W(x)<E$ for $y_i<x<z_i$.
The index $i=1,2$ refers to two different types of the potential minima
(lower and upper).

For large $W_0 \gg \D W$,
the lowest energy band is narrow and lies well below
the potential maximum $W_0$.
It implies small tunneling probability $T(E)$ and small $\D \Phi(E)$
which are both relatively insensitive to $E$.
The average phase can be approximated by a linear function
$\Phi(E)\approx \alpha (E-E_0)$ around the value $\pi$
[first minimum of $\cos^2(\Phi/2)$] which leads to the effective
universal relation
\begin{equation}
  E(p)-E_0 \approx \pm\sqrt{M^2+J^2\cos^2(\a p)}
\end{equation}
where $M:=(1-T)\, \sin^2(\Delta \Phi)/\alpha^2$ and $J^2:=T/\alpha^2$.
The approximation holds uniformly for all $p\in(-\pi/2\a,\pi/2\a)$
as long as $\D W \ll W_0$ (for more details, see \cite{Bands}).
%%\footnote{The approximation accuracy is comparable with the
%\textit{single cosine} approximation \eqref{E-1cos} in the original
%periodic potential which is widely used.}
%to the exact dispersion relation $E_{2\pm}(p)$.
Using again the WKB approximation, we can estimate the parameters
\begin{equation}
\label{WKB-J}
J \approx
\frac{4}{\pi}\,\sqrt{W_0 E_R}\,
\exp\left\{-\frac{\pi}{4}\sqrt{\frac{W_0}{E_R}}\right\}, \qquad
M \approx \frac{\Delta W}{2}
\end{equation}
where $E_R=k^2/(2M_\text{atom})=\pi^2/(8M_\text{atom} a^2)$ is the recoil
energy and $M_\text{atom}$ the mass of the atoms moving in the
doubly-periodic potential $W(x)$.

%%%%%%%%%%%%%%%%%%%%%%%%%%%%%%%%%%%%%%%%%%%%%%%%%%%%%%%%%%%%%%%%%%%%%%%%%%%%%%
%%%%%%
\subsection{Wannier functions and sites}

Let us discuss the transition from the simply periodic potential to the
doubly periodic one on the level of the associated Hamiltonian.
Starting with the single-particle Schr{\"o}dinger Hamiltonian
describing atoms in an external potential
\bea
\h H = -\frac{1}{2 M_{\text{atom}}}\nabla^2 + W(x)
\ea
and performing the standard steps we obtain, for the original periodic
potential $W(x)=W_0 \sin^2(2kx)$, the usual second-quantized Hamiltonian
in momentum space
\begin{equation}
\h H_{\rm original}
=
\frac{\a}{2\pi}\int_{-\pi/\a}^{\pi/\a} dp\,E_{\rm original}(p)\,
\h\psi\s(p) \h\psi(p)
\,,
\end{equation}
where we consider the lowest band only with
$E_{\rm original}(p)\approx - J \cos(\a p)$
%
%On the level of the Schr{\"o}dinger Hamiltonian
%$\h H_1 = -\frac{1}{2 M_{\text{atom}}}\D + W_1(x)$,
%describing atoms in the periodic potential $W_1(x)=W_0 \sin^2(2kx)$,
%the following transition takes place. In the momentum space we have
%\begin{equation}
%  \h H_1 = \frac1{2\pi}\int_{-\pi/\a}^{\pi/\a} E_1(p)\,
%\h\psi\s(p) \h\psi(p)\, dp
%\end{equation}
%with $E_1(p)\approx - J \cos(\a p)$
%
(for convenience we shifted the energy scale by the constant $E_0$
what has no physical consequences).
After switching on the doubly-periodic perturbation $\D W \sin^2(kx)$,
the energy spectrum undergoes a transition
$E_{\rm original}(p)\ra E(p)
\approx \pm\sqrt{M^2+J^2\cos^2(\a p)}$ and the Hamiltonian becomes
\begin{equation}
  \h H  = \intpia dp\,
  %\left(\begin{array}{c}
  \left(\begin{array}{c}
     \h\chi(p) \\ \h\psi(p)
  \end{array}\right)^\dagger %\s
  %\end{array}\right)\s
  {\cal M}
  %\left(\begin{array}{c}
  \left(\begin{array}{c}
    \h\chi(p) \\ \h\psi(p)
  \end{array}\right)
  %\end{array}\right)
\end{equation}
with $\h\chi(p):=\h\psi(p+\pi)$
[assuming periodicity $\h\psi(p+2\pi) = \h\psi(p)$] and
\begin{equation}
  {\cal M} =
    %\left(\begin{array}{c}
    \left(\begin{array}{cc}
     \sqrt{M^2+J^2\cos^2(\a p)} & 0 \\
     0 & -\sqrt{M^2+J^2\cos^2(\a p)}
  \end{array}\right).
  %\end{array}\right).
\end{equation}
This Hamiltonian has the same form as the one for the discretized
Dirac equation (\ref{H0-a-b-sqrt}) when we set $J=1/\a$.

For the two separated energy bands there exist two separate sets of
Wannier functions on the lattice: the ``lower'' and the ``upper''
centered at even and odd sites, respectively.
But these Wannier functions turn out to be poorly localized on the lattice
({somewhat} analogously to continuous quantum field theory where free
particles with fixed energy are not localized in space).
In order to achieve optimal localization it is preferred to switch to the
set of operators introduced already in \eqref{diag-a(p)}-\eqref{U(p)}
\begin{equation}
  %\left(\begin{array}{c}
  \left(\begin{array}{c}
    \h a(p) \\ \h b(p)
  \end{array}\right)
  %\end{array}\right)
  = U(p)\s
  %\left(\begin{array}{c}
  \left(\begin{array}{c}
    \h\chi(p) \\ \h\psi(p)
  \end{array}\right)
  %\end{array}\right)
\end{equation}
what transforms the matrix ${\cal M}$ via the similarity transformation
${\cal M}'=U\s {\cal M} U$ to
\begin{equation}
  {\cal M}' =
  %\left(\begin{array}{c}
  \left(\begin{array}{cc}
     M & J\cos(\a p) \\
     J \cos(\a p) & -M
  \end{array}\right).
  %\end{array}\right).
\end{equation}
Now, going from the momentum to the site representation by inverting the
Fourier transformation \eqref{a-b-Fourier}
\begin{equation}
\h a_{2n} = \intpia dp\,e^{2in\a p} \h a(p)\,,
\qquad
\h b_{2n+1} = \intpia dp\, e^{i(2n+1)\a p} \h b(p)\,,
\end{equation}
we obtain the free Hubbard Hamiltonian \eqref{H-lattice-a-b} with $J=1/(2\a)$.
By this construction $\h a\s_{2n}$ and $\h b\s_{2n+1}$ create two types of
particles in two types of Wannier states exponentially localized at even and
odd sites, respectively.
%The Wannier functions can be obtained from the inverse Fourier transformations
%\begin{eqnarray}
%   a_{2n} &= \frac{1}{\pi} \int_0^\pi e^{2ikp}  a(p) dp, &
%   b_{2n+1} &= \frac{1}{\pi} \int_0^\pi e^{i(2n+1)p}  b(p) dp
%\end{eqnarray}
%where now the $a_n, b_n, a(p), b(p)$ (with no hat) refer to the
%corresponding wave functions.
However, they do not give rise to ``positive'' and ``negative energy sites''
as they mix energies from both bands.
{This} can be best seen in the limiting case $M \ll J$ where the
Wannier functions [up to terms $\Ocal{M/J}$]
\begin{equation}
   a_{2n} \cong \frac1{\sqrt{2}} (\chi_{2n} - \psi_{2n}), \qquad
   b_{2n+1} \cong \frac1{\sqrt{2}} (\chi_{2n+1} + \psi_{2n+1})
\end{equation}
are build from the difference and sum of the single-band Wannier functions
for the lower and upper bands defined as
\begin{equation}
\psi_n := \intpia dp\, e^{in\a p}  \psi(p) \,,
\qquad
\chi_n := \intpia dp\, e^{in\a p}  \chi(p) \,.
\end{equation}

%\myboxm{PLOT or SKETCH: Wannier functions\\[10em]}

\subsection{Physical parameters}

In order to discuss the experimental realizability of our quantum simulator,
let us summarize the conditions on the involved parameters.
Strictly speaking, the WKB approximation used above requires $W_0\gg E_R$
which then implies $E_R\gg J$ via Eq.~(\ref{WKB-J}).
However, even if we relax these conditions to
\bea
W_0 > E_R > J
\,,
\ea
we still get qualitatively the same picture.
What is crucial, however, is the applicability of the single-band
Fermi-Hubbard Hamiltonian~ (\ref{Fermi-Hubbard0}).
To ensure this, we demand that the local oscillator frequency
$\omega_{\text osc}$ in the potential minima be much smaller than $J$.
In addition, the continuum limit -- i.e., that the discretized
expression~(\ref{continuum}) provides a good approximation --
requires $J \gg M$, i.e., $1/\a\gg M$.
For the same reason, the change $\Delta\Phi_n=\Phi_{n+1}-\Phi_n$ of
the analogue of the electrostatic potential $\Phi_n$ from one site
to the next should be smaller than $M$.
Over many sites, however, this change can well exceed the mass
gap $2M$, which is basically one of the conditions for the
Sauter-Schwinger effect to occur.
Finally, the effective temperature $T$ should be well below the
mass gap $2M$ in order to avoid thermal excitations.
In summary, the analogue of the $e^+e^-$ pair creation can be simulated
if the involved scales obey the hierarchy
\begin{eqnarray}
\omega_{\text osc} \gg J \gg M \gg T
\,.
\end{eqnarray}
Let us insert some example parameters.
The recoil energy $E_R$ of ${}^6$Li atoms in an optical lattice made of
light with a wavelength of 500~nm is around $E_R\approx7\,\mu\text K$.
Thus, if we adjust the potential strength to be $W_0=10\,\mu\text K$,
the hopping rate $J$ would be around $5\,\mu\text K$ which is still
sufficiently below the local oscillator frequency $\omega_{\text osc}$
of around $34\,\mu\text K$.
Then a perturbation of $\Delta W=1\,\mu\text K$ created by light with a
wavelength of 1000~nm would induce an effective mass $M$ of 500~nK
and thus the effective temperature should be below that value --
which is not beyond present experimental capabilities.

\black

\xnewpage
\section{Spontaneous pair creation on the lattice}

%%%%%%%%%%%%%%%%%%%%%%%%%%%%%%%%%%%%%%%%%%%%%%%%%%%%%%%%%%%%%%%%%%%%%%%%%%%%%%%

\begin{figure}[h]
\begin{center}
  \includegraphics[width=1\linewidth]{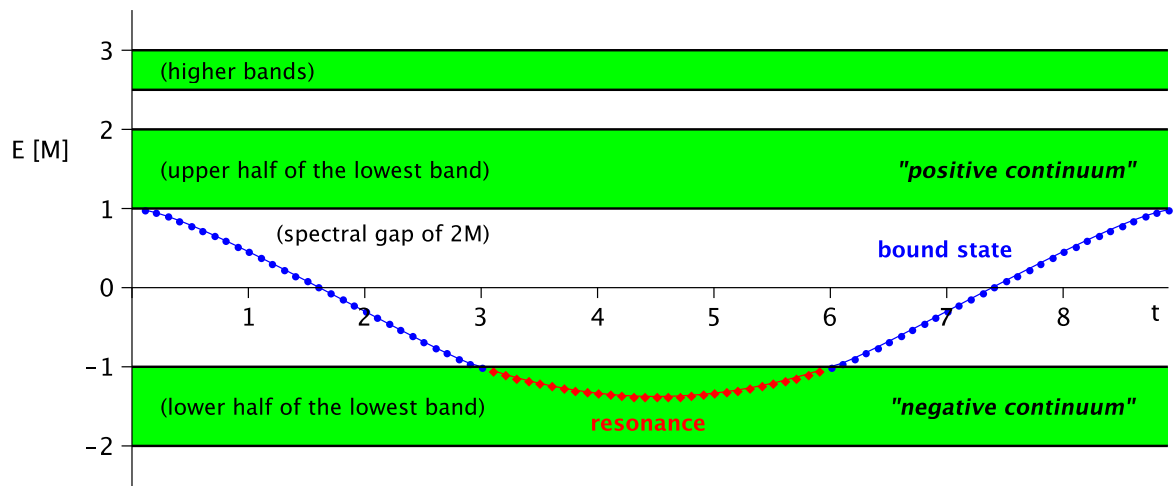}
\end{center}
\caption{Time-dependent bound state/resonance interpolating between two
halves of the lowest energy band after splitting it by a doubly-periodic
perturbation of the optical potential. (Bands and gap not to scale.)}
\label{fig:timedep-bs}
\end{figure}

The above established analogy between the (discretized) second quantized
Dirac field describing electrons and positrons in an electric field, on the
one hand, and the (Fermi or Bose, see Sec. \ref{sec:BF-mapping}) Hubbard
model describing ultra-cold atoms in an optical lattice, on the other hand,
enables laboratory simulations of some of the relativistic phenomena of
strong-field QED.
The original {Sauter-Schwinger} effect \cite{Schwinger} with a constant
electric field $E$ would correspond to a static tilted optical lattice
with $\Phi(x) = E x$ (the so called Wannier-Stark ladder,
see e.g. \cite{Wannier-Stark}).
For nonzero values of $E>0$ one would expect a constant {rate} of
spontaneously generated particles and holes
(cf. formula \eqref{P-Schwinger}), depending non-perturbatively on $E$.
{
However, a constant electric field $E$ is unrealistic from an
experimental point of view.
An electric field which is localized in space and time is simpler to
handle both, experimentally and conceptually
(see e.g. \cite{Sauter, BongRuij}).
Therefore, in the following we discuss the process of
\textit{analogue spontaneous pair creation} in presence of an external
localized potential which will be slowly switched on to a supercritical
value -- admitting one bound state to dive into the negative continuum --
and then switched off, as discussed in Sec \ref{sec:SpPCr}.
}
%
%such setting might be experimentally problematic due to strong
%finite-size effects on the lattice.
%
%An electric field which is localized in space and time is simpler to
%handle both, in physical realization and in mathematical calculations
%(see e.g. \cite{Sauter, BongRuij}).
%
%Therefore, we will further discuss the process of
%\textit{analogue spontaneous pair creation} in presence of an external
%localized potential which will be slowly switched on to a supercritical
%value and then switched off, as discussed in Sec \ref{sec:SpPCr}.
%
In presence of the attractive potential a bound state will form in the gap
$2M$ between the two lowest bands (formed from the lowest band splitted
due to the doubly-periodic perturbation). During the time-dependent
process, the bound state will slowly reach the lower band and then turn
into a resonance lying within this band (see Fig. \ref{fig:timedep-bs}).
The resonance will then decay causing an instability of the Fermi state
(our \textit{analogue vacuum state}) which will
spontaneously decay to an energetically more favorable state with a
particle-hole pair present.
The ``particle'' (an atom excited above the Fermi level) will stay bound
by the attractive potential while the ``antiparticle''
(hole in the Fermi sea) will {be in a scattering state}.
%
%get scattered.
%
%

%%%%%%%%%%%%%%%%%%%%%%%%%%%%%%%%%%%%%%%%%%%%%%%%%%%%%%%%%%%%%%%%%%%%%%%%%%%
\subsection{Experimental procedure}

\begin{figure}[h]
\begin{minipage}{0.67\linewidth}
  \begin{center}
  \includegraphics[width=1\linewidth,height=0.1\textheight]{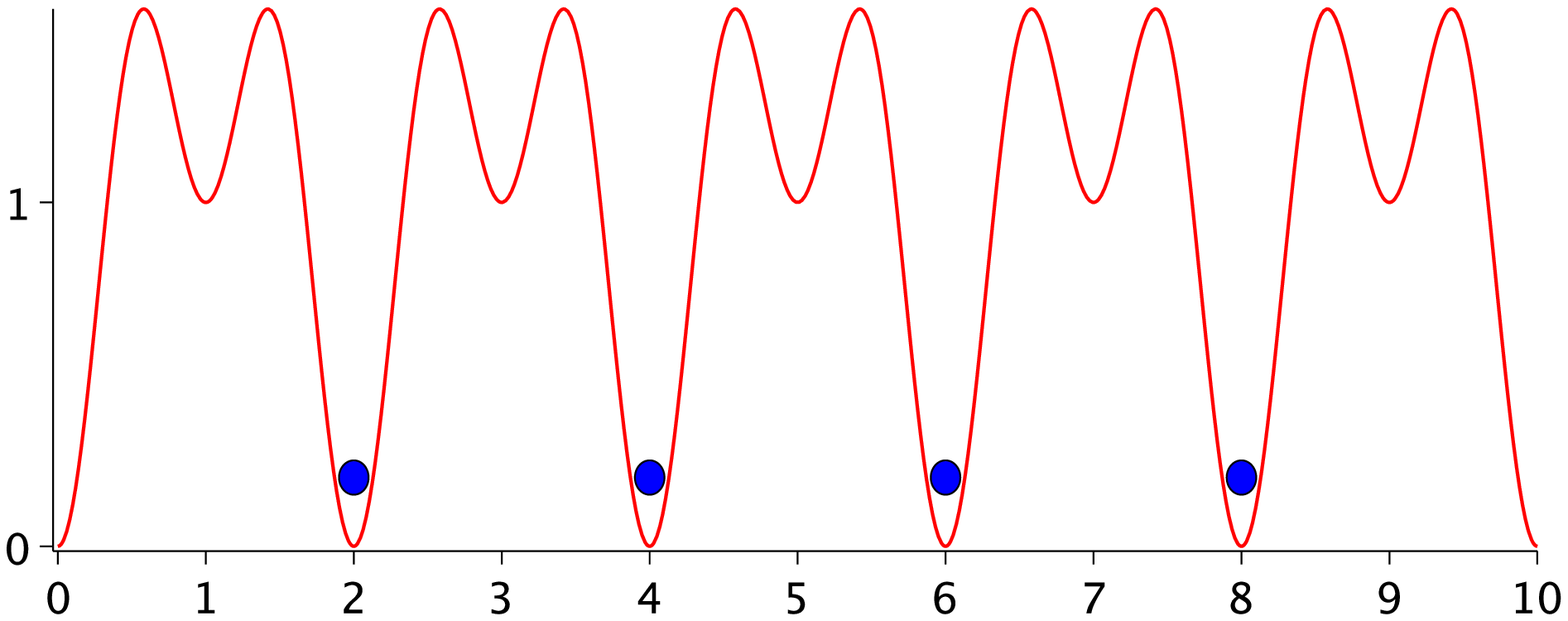}
  \includegraphics[width=1\linewidth,height=0.1\textheight]{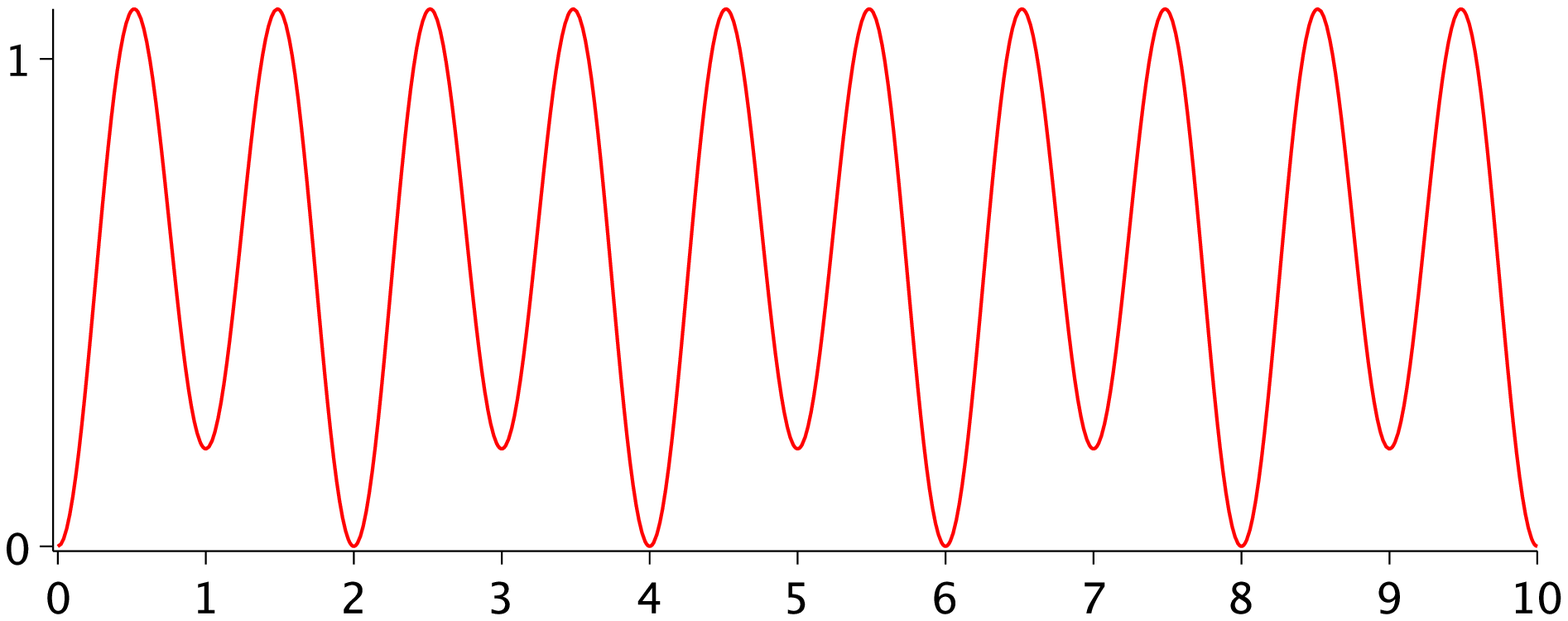}
  \includegraphics[width=1\linewidth,height=0.15\textheight]{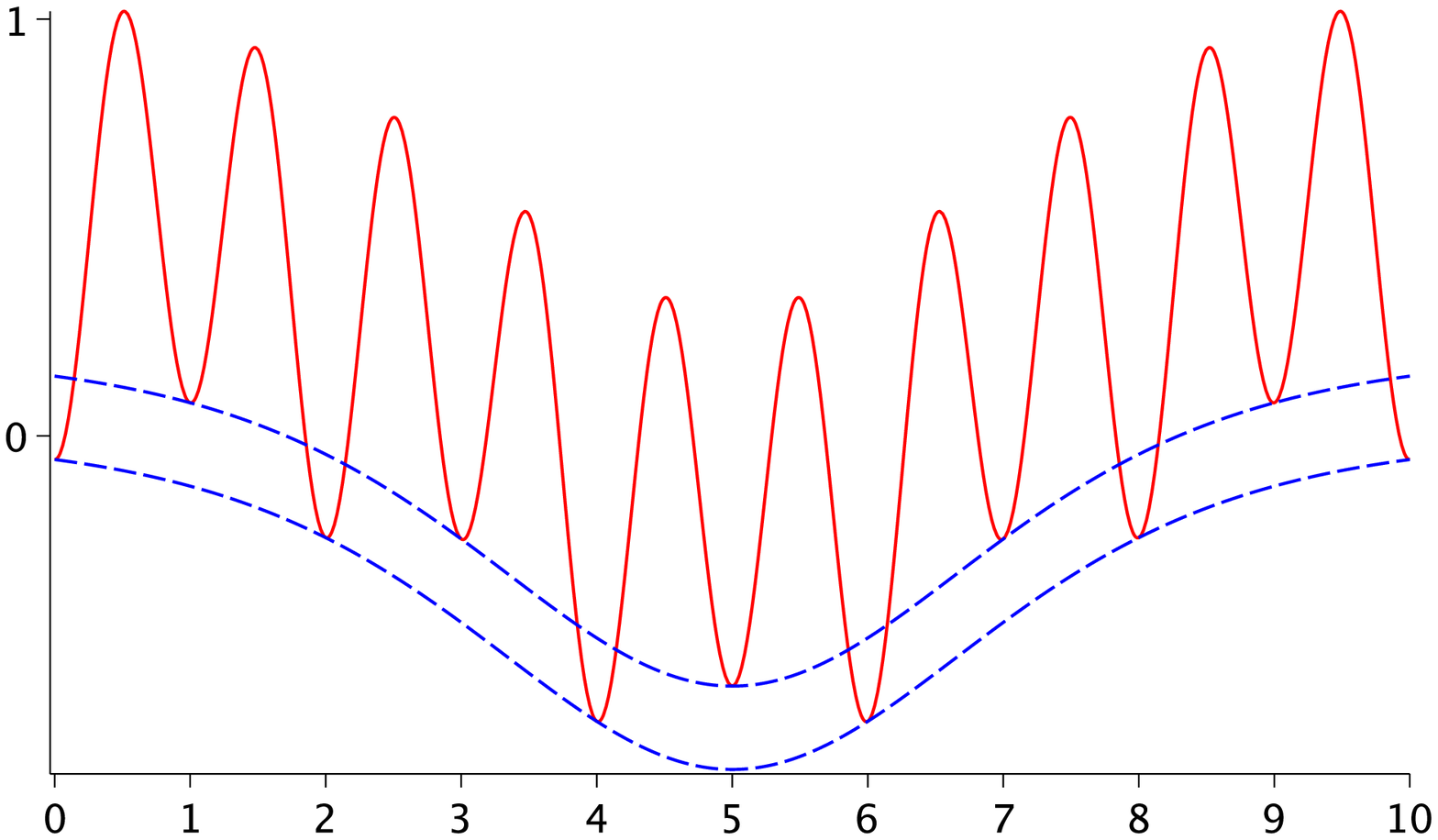}
  \includegraphics[width=1\linewidth,height=0.1\textheight]{dppot-low}
  \includegraphics[width=1\linewidth,height=0.1\textheight]{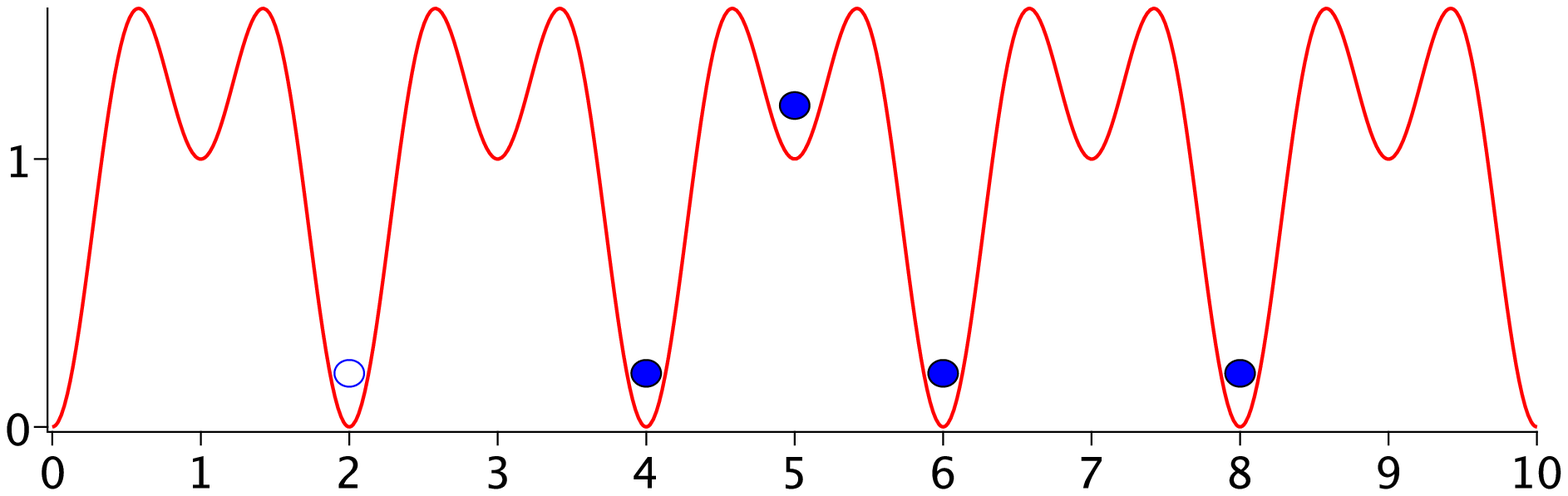}
  \end{center}
\end{minipage}
\hfill
\begin{minipage}{0.33\linewidth}
 \includegraphics[width=1\linewidth]{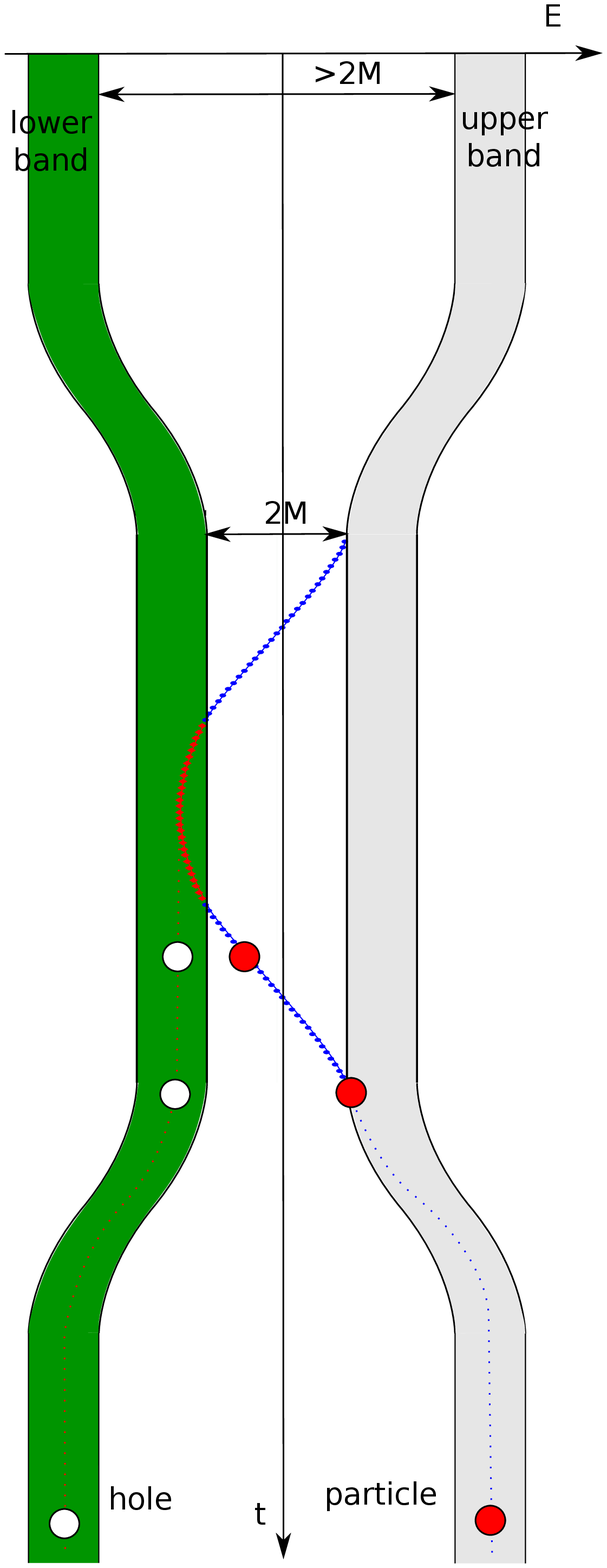}
\end{minipage}
\caption{Sketch (not to scale) of the stages of the simulation
(from top to bottom).
Left: The solid red curves represent the sum of the optical lattice
potential $W(x)$ and the dynamically switched supercritical potential
$\Phi(x)$ as a function of position $x$ %versus the lattice spacing
while the dashed blue curves correspond to the effective electric
potential $\Phi$.
The blue solid dots are particles and the empty circle is a hole.
Right: Band structure and a bound state changing in time.
}
\label{Fig:stages}
\end{figure}

Simulation of the ``pair production'' in the optical lattice requires
preparation of the initial quantum state of the atoms corresponding
to the \textit{Dirac sea} in QED.
This can be achieved by keeping a large value of $\Delta W$
(separation of two bands by a large gap) during the
cooling phase and generating a clear Fermi level at half filling of the
lattice (all particles in the lower sites, see Fig. \ref{Fig:stages}, top).
Next, the value of $\Delta W$ should be adiabatically decreased to values
well below $W_0$ to achieve $M\ll J$ and allow the atoms for
\textit{dispersion} across the lattice.
The atoms become delocalized but still the lower band is fully filled while
the upper band remains empty (second picture in Fig.~\ref{Fig:stages}).
Then the ``external'' potential $\Phi$ (which mimics the electric potential)
can be slowly switched on to reach a supercritical value.
In that phase, the ground  state (``vacuum'') will get rearranged via
tunneling from the lower band to the upper band
(analogue of the Sauter-Schwinger effect, third picture in
Fig.~\ref{Fig:stages}).
Such an created particle-hole pair will tend to separate on the lattice so
that when the potential is slowly switched off after some delay the pair
will not be able to annihilate any more
(fourth picture in Fig.~\ref{Fig:stages}).
That mimics the well known \textit{spontaneous pair creation} known from QED.
Finally, in order to detect the ``pair'' in experiment, the value of
$\Delta W$ can be {adiabatically} increased again thus leading to
energetic separation of the created particle and hole represented by an
atom in one of the upper minima and a missing atom in one of the lower
minima (fifth picture in Fig.~\ref{Fig:stages}).
The atom in one of the upper sites and the hole (missing atom) in
one of the lower minima could be detected via site-resolved imaging
\cite{Bloch-SingleAtomResolvedFluorescence}.
Another option could be blue-sideband-detuned optical transitions which
are resonant to the oscillation frequency $\omega_{osc}$ of the
upper minima but not to those in the lower sites.

%Detection of an atom in the upper site should not present a difficulty,
%for instance, by absorption of photons tuned to $\omega_{osc}$
%(Rabi oscillations) which
%is different for the upper and lower sites in the final setting or by
%site-resolved imaging \cite{Bloch-SingleAtomResolvedFluorescence}.

\black

%%%%%%%%%%%%%%%%%%%%%%%%%%%%%%%%%%%%%%%%%%%%%%%%%%%%%%%%%%%%%%%%%%%%%%%%%%%%%
%%%%%%%%%%%%%%%
\subsection{Bose-Fermi mapping} \label{sec:BF-mapping}
Since it is typically easier to cool down bosonic than fermionic atoms,
let us discuss an alternative realization based on bosons in an optical
lattice.
To this end, we start with the Bose-Hubbard Hamiltonian
\begin{equation}
\label{Bose-Hubbard}
\h H = -J \sum_{n}
\left[\h d\s_{n+1} \h d_n + \h d\s_{n} \h d_{n+1}\right] +
\sum_n V_n \h d\s_n \h d_n +
\frac{U}{2}\sum_n (\hat d^\dagger_n)^2\,\hat d_n^2
\,,
\end{equation}
which has the same form as the Fermi-Hubbard Hamiltonian
\eqref{H-lattice-a-b} after replacing the fermionic $\hat c_n$ by bosonic
$\hat d_n$ operators, but with an additional on-site repulsion term $U$.
For large $U\gg J$ (which can be controlled by an external magnetic field
via a Feshbach resonance, for example), we obtain the bosonic analogue
of ``Pauli blocking'', i.e., at most one particle can occupy each site
{$\hat d_n^2\ket{\Psi}\approx0$}.
Neglecting all states with double or higher occupancy, we can map these
bosons exactly onto fermions in one spatial dimension via
\begin{equation}
\h d_n = \exp\left(-i\pi\sum_{m<n} \h c\s_m \h c_m\right) \h c_n
\,.
\end{equation}
Via this transformation, the bosonic commutation relations
$[\hat d_n,\hat d^\dagger_m]=\delta_{nm}$ and
$[\hat d^\dagger_n,\hat d^\dagger_m]=[\hat d_n,\hat d_m]=0$
are exactly mapped onto the fermionic anti-commutation relations
$\{\hat c_n,\hat c^\dagger_m\}=\delta_{nm}$ and
$\{\hat c_n,\hat c_m\}=\{\hat c^\dagger_n,\hat c^\dagger_m\}=0$.
As a result, we obtain the same physics as described by the
Fermi-Hubbard Hamiltonian \eqref{H-lattice-a-b}.

\black

%%%%%%%%%%%%%%%%%%%%%%%%%%%%%%%%%%%%%%%%%%%%%%%%%%%%%%%%%%%%%%%%%%%%%%%%%%
\subsection{Interactions}

Apart from investigating the pair creation probability for space-time
dependent electric fields $E(t,x)$, this quantum simulator for the
Sauter-Schwinger effect could provide some insight into the impact of
interactions.
For example, including dipolar interactions of the atoms, we would get the
coupling Hamiltonian $D_{nm}\h c\s_{n}\h c\s_{m} \h c_{n}\h c_{m}$ with
$D_{nm}\propto|n-m|^{-3}$.
As an example for permanent dipole moments, we may consider ${}^{52}$Cr
atoms possessing a rather large magnetic moment.
However, the associated interaction energy $D_{nm}$ would be below one
nano-Kelvin and thus probably too small to generate significant effects.
Therefore, let us consider externally induced dipole moments.
For example, ${}^6$Li atoms can be electrically polarized by an external
electric field of order $10^8~\text V/m$
(which can be realized experimentally) such that the induced electric
dipole moments generate interaction energies up to a few $\mu$K.
\black
%
%The interaction energy $D_{nm}$ due to the large magnetic
%moments of ${}^{52}$Cr atoms would be below one nano-Kelvin.
%
%Larger interaction energies up to a few $\mu$K can be obtained for
%${}^6$Li atoms by external electric fields of order $10^8~\text V/m$
%which induce electric dipole moments \cite{dipole-dipole}.
%
By aligning the atomic dipole moments parallel or perpendicular to the
lattice, we may even switch between attractive $D_{nm}<0$ and repulsive
$D_{nm}>0$ interactions.
Note that this goes far beyond the simulation of the classical
Dirac equation and requires the full quantum many-particle
Hamiltonian.

%\mybox{Dipole-dipole in Rydberg atoms ?}

%%%%%%%%%%%%%%%%%%%%%%%%%%%%%%%%%%%%%%%%%%%%%%%%%%%%%%%%%%%%%%%%%%%%%%%%%%
%%%%%%%%%%%%%%%%%%
\xnewpage
\appendix
\section{Potential localized at one
site (delta-like)}

\begin{figure}
  \begin{center}
\includegraphics[width=0.5\linewidth,height=0.2\textheight]{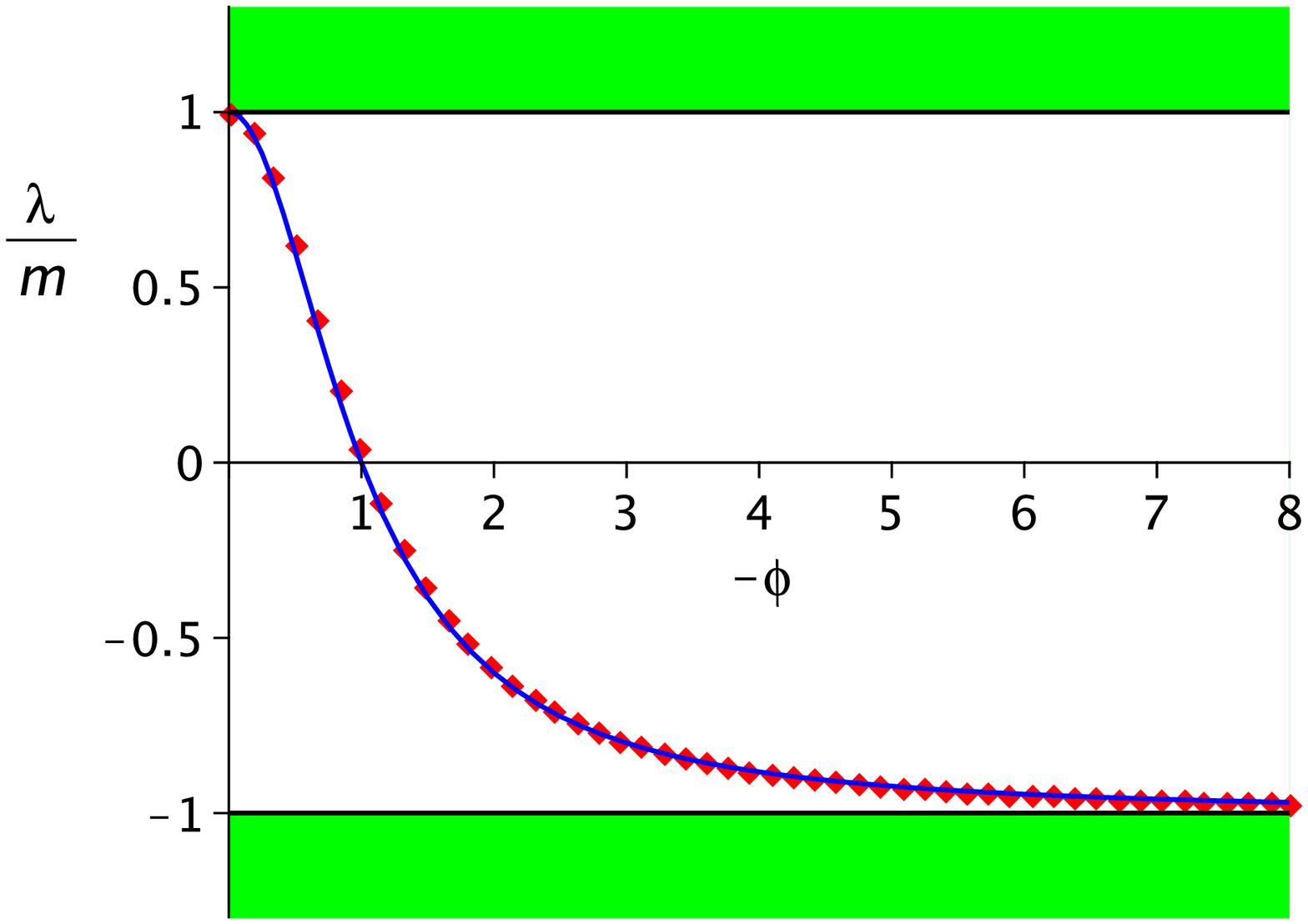} \hfill
\includegraphics[width=0.45\linewidth,height=0.2\textheight]{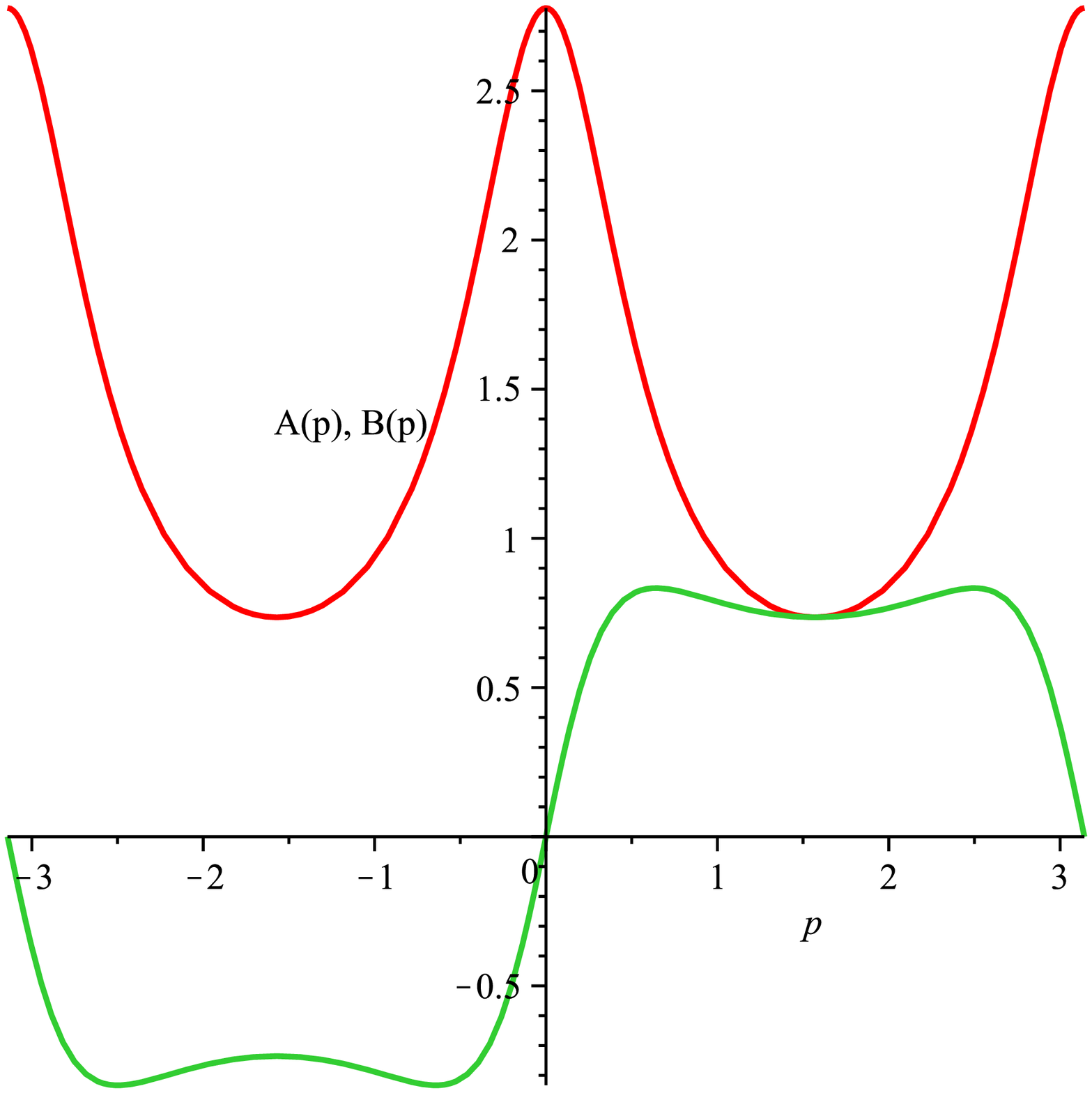}
\caption{Left: The eigenvalue $\la/M$ as a function of the potential's
strength $\phi$ for the discretized Dirac equation (red solid line) compared
to the continuous Dirac equation with the delta potential (blue dashed line).\\
    Right: The momentum distribution $A(p), B(p)$ for a bound state with
$\la=0.8 M$.}\label{fig:Ex:la}
  \end{center}
\end{figure}

In order to compare the discretized Dirac Hamiltonian with its
continuum version, we consider an example where both can be solved
analytically.
This is possible for a Dirac delta like potential localized
at one lattice site
\begin{equation}
  \phi_n=\phi\, \frac{\delta_{0,n}}{\a} \qquad \text{with} \qquad \phi(p) = \phi
\end{equation}
which corresponds to $\phi(x)=\phi\, \delta(x)$ in the continuous case $\a\ra 0$.
It has the advantage that a closed formula for the bound-state energy
can be found analytically.
The Hubbard Hamiltonian \eqref{H0-a-b-cos} with added potential reads then
\begin{eqnarray}
%\begin{split}
\fl
\h H &=& \intpia dp\,
\left[ M \left( \h a(p)\s \h a(p) - \h b(p)\s \h b(p) \right)
+ \J \cos(\a p) \left( \h a(p)\s \h b(p) + \h b(p)\s \h a(p) \right) \right] +
%\right.
%\nn
%     &
%\left.\phantom{\intpi\big(}
%+ \J \cos(\a p) \h a(p)\s \h b(p) + \J \cos(\a p) \h b(p)\s \h a(p)\right] +
\nn
\fl
&& + \frac{\phi\, \a^2}{\pi^2}
\left[
\int_{-\pi/2\a}^{\pi/2\a} dp_1\,\h a^\dagger(p_1)
\;
\int_{-\pi/2\a}^{\pi/2\a} dp_2\,\h a(p_2)
+
\int_{-\pi/2\a}^{\pi/2\a} dp_1\,\h b^\dagger(p_1)
\;
\int_{-\pi/2\a}^{\pi/2\a} dp_2\,\h b(p_2)
\right]
\,,
%
%\intpi \h a(p_1)\s dp_1\cdot \intpi \h a(p_2) dp_2
%\right.
%\nn
%\left.\phantom{\phi\big(}
%+ \intpi \h b(p_1)\s dp_1\cdot \intpi \h b(p_2) dp_2 \right]
%\end{split}
\end{eqnarray}
where we set $J=1/\a$.
%
%(This can always be done without restriction of generality.)
%
Obviously, it satisfies $\h H \vac = 0$ for the vacuum vector $\vac$
such that $\h a(p) \vac = \h b(p) \vac = 0$.
However, we want to find a one-particle eigenstate $\v{\chi}$ of the
Hamiltonian $\h H$ with eigenvalue $\la$ lying in the gap $-M<\la<M$
(which corresponds to a bound state for the discretized Dirac equation).
Then we want to study the dependence of the eigenvalue $\la$ on the
strength of the potential $\phi$.
The general one-particle state can be written as
\begin{equation}
\v\chi
=
\left[ \intpia dp\, A(p) \h a\s(p) + \intpia dp\, B(p) \h b\s(p) dp\right] \vac
\end{equation}
with two complex functions $A(p)$ and $B(p)$.
By projecting the eigenvalue equation $\h H \v\chi = \lambda \v\chi$
on the states $\cav \h a(p)$ and $\cav \h b(p)$ we arrive at a system of
equations
\begin{equation}
\label{A(p)-B(p)=A-B}
\underbrace{\left(\begin{array}{cc}
    M-\la, & \J \cos(\a p) \\
    \J \cos(\a p), & -M-\la
  \end{array}\right)}_{=:\M}
  \left(\begin{array}{c}
    A(p)\\
    B(p)
  \end{array}\right)
=
  - \phi\,
\intpia dp
  \left(\begin{array}{c}
     A(p)  \\
     B(p)
  \end{array}\right). % =
%  -\phi
%  \left(\begin{array}{c}
%    \bar A \\
%    \bar B
%  \end{array}\right).
\end{equation}
Integration of both sides over $p$ leads to the relations
\begin{eqnarray}
  \bar A &:= \intpia A(p) dp = \frac{\intpia dp\, \J \cos(\a p) B(p)}{\la-M-\phi}\,,
\nn
  \bar B &:= \intpia B(p) dp = \frac{\intpia dp\, \J \cos(\a p) A(p)}{\la-M-\phi}\,.
\end{eqnarray}
Observe that the right-hand side of \eqref{A(p)-B(p)=A-B} is equal to
$-\phi (\bar A, \bar B)^T$.
We can invert the matrix $\M$, whose determinant
$\det \M=\la^2-M^2-\J \cos^2(\a p)<0$ never vanishes, to obtain
\begin{equation} \label{Ex:A-B-sys}
  \left(\begin{array}{c}
    A(p)\\
    B(p)
  \end{array}\right) =
  \frac{\phi}{\la^2-M^2-\J \cos^2(\a p)}
  \left(\begin{array}{cc}
    \la+M, & \J \cos(\a p) \\
    \J \cos(\a p), & \la-M
  \end{array}\right)
  \left(\begin{array}{c}
    \bar A \\
    \bar B
  \end{array}\right).
\end{equation}
Integrating both sides over $p$ again and using
\begin{eqnarray} \label{Ex:Abar-Bbar}
  \intpia \frac{dp}{\la^2-M^2-\J \cos^2(\a p)}
&= -\frac1{\sqrt{M^2-\la^2}\sqrt{M^2+\J^2-\la^2}}
\,,
\nn
  \intpia \frac{\J \cos(\a p)\ dp}{\la^2-M^2-\J \cos^2(\a p)}
&= 0
\,,
\end{eqnarray}
we obtain consistency conditions
\begin{eqnarray}
  \left(\begin{array}{c}
    \bar A\\
    \bar B
  \end{array}\right) =
  \frac{-\phi}{\sqrt{M^2-\la^2}\sqrt{M^2+\J^2-\la^2}}
  \left(\begin{array}{c}
    (\la+M) \bar A \\
    (\la-M) \bar B
  \end{array}\right)
\end{eqnarray}
which cannot be satisfied at the same time except when (at least)
one of the constants $\bar A, \bar B$ vanishes.
Assume first, it is $\bar B=0$ (the case $\bar A=0$ will be discussed below).
Then we need to solve the algebraic equation
\begin{equation}
  -\phi \frac{\la+M}{\sqrt{M^2-\la^2}\sqrt{M^2+\J^2-\la^2}} = 1
\end{equation}
for the function $\la(\phi)$.
It has, in general, three solutions what can be better seen from
\begin{equation}
  (M-\la)(M^2+\J^2-\la^2)=\phi^2(M+\la)
\,.
\end{equation}
{For $\phi=0$, these three solutions start from the points
$M, \pm\sqrt{M^2+\J^2}$, i.e., the edges of the continuous spectrum.}
We are interested in the perturbations of the eigenvalue $\la=M$ for
negative values of $\phi$, i.e. for a bound state separating from the
bottom of the upper band.
{For small $\phi$, $\la(\phi)$ behaves like
$\la(\phi)\cong M (1 - 2 \phi^2)$.}
For increasing values of $|\phi|$ it monotonically decreases to
$-M$ but never reaches this value
(more precisely, $\la(\phi) \cong -M + 2M/\phi^2$ for $|\phi| \gg 1$)
-- see Fig. \ref{fig:Ex:la}.
It shows that there is no \textit{supercriticality} in this potential, i.e.,
crossing of the value $\la(\phi)=-M$ at finite $\phi$, what is analogous
to the continuous case $\phi(x)=\phi\,\delta(x)$ in which
$\la(\phi) = M (1-\phi^2)/(1+\phi^2)$ \cite{Dirac+Delta1D}.

From equation \eqref{Ex:A-B-sys} we can also find the eigenvector to the
eigenvalue $\la$ %(with $\bar B=0$)
\begin{equation}
  \left(\begin{array}{c}
    A(p)\\
    B(p)
  \end{array}\right) =
  \frac{\phi\ \bar A}{\la^2-M^2-\J \cos^2(\a p)}
  \left(\begin{array}{c}
    \la+M \\
    \J \cos(\a p)
  \end{array}\right)
\end{equation}
where the value of $\bar A$ is to be determined from the normalization
condition
\begin{equation}
  \intpi dp\,\left( |A(p)|^2+|B(p)|^2\right) = 1
\,.
\end{equation}
Since both $|A(p)|^2$ and $|B(p)|^2$ are even functions of $p$,
we have
\begin{equation}
  \intpi dp\,\left( |A(p)|^2+|B(p)|^2\right) \J \cos(\a p) = 0
\,,
\end{equation}
which reflects the fact that the (discretized) momentum {$id/dx \ra \J \cos(\a p)$} vanishes in the
bound state.
Note also that the condition $\bar{B}=0$ implies $B_0=0$ what means that
the ``antiparticles'' are repelled from the site at which the potential
is localized.

The case $\bar A=0$ is fully analogous and the solutions can be obtained
by a symmetry transformation: $\la\ra-\la$ and $\phi\ra-\phi$.
The bound state emerges then from the lower band (negative``continuum'')
at $\la=-M$ and goes up for positive potentials $\phi$.

\paragraph{Remark 1.} The considered potential is localized at
\textit{one site} $n=0$ in that sense that it interacts with both
types of particles via $a_0$ and $b_0$.
But in fact, $a_0$ and $b_0$ are \textit{two different} (neighbouring)
sites brought to $k=0$ by a convenient renumbering.
It is also possible to consider the potential to be localized in such a
way that it interacts with only one type of particles, say via $a_0$.
Then the equations get slightly modified (the term $\bar{B}$ disappears
at some places) and we obtain $\bar{B}=0$ as consequence of
\eqref{Ex:A-B-sys} and \eqref{Ex:Abar-Bbar}.
From that point on, the solution is identically the same to the previous case.
It means that it plays no role whether we consider potentials
localized at one site interacting with only one or with both types of
particles because, in the latter case, the solutions split into two
symmetric cases of the former type.

\paragraph{Remark 2.} The above method of calculating $\la(\phi)$ works
only for potentials for which $\intpi dp\,\phi(p-q) \det \M(p)^{-1}$
%and $\intpi V(p-q) \det M(p)^{-1} dp$ are
is independent of $q$, i.e. for $\phi(p)=const$ which corresponds to
$\phi_n \sim \delta_{n,0}$. Unfortunately, it cannot be generalized
to more complex potentials in a simple way.

%%%%%%%%%%%%%%%%%%%%%%%%%%%%%%%%%%%%%%%%%%%%%%%%%%%%%%%%%%%%%%%%%%%%%%%%%%%%%%%%%%%%%%%%%%%%
%\xnewpage
%\subsection{Example: Supercritical potential (lattice)}

%\mybox{Is there a potential analytically solvable on the lattice?!}

%%%%%%%%%%%%%%%%%%%%%%%%%%%%%%%%%%%%%%%%%%%%%%%%%%%%%%%%%%%%%%%%%%%%%%%%%%%%%%%
% \bibliography{basename of .bib file}
%\bibliography{lattices}
%\bibliographystyle{unsrt}
%\bibliographystyle{apsrev}
%\xnewpage

\end{document}